\documentclass[conference]{IEEEtran}
\IEEEoverridecommandlockouts
\usepackage{cite}
\usepackage{amsmath,amssymb,amsfonts}
\usepackage{graphicx}
\usepackage{textcomp}
\usepackage{xcolor}
\usepackage{mathtools}
\usepackage{float}
\usepackage{subfig}
\usepackage{graphicx}
\usepackage{textcomp}
\usepackage{xcolor}
\usepackage{listings}
\usepackage{xspace}
\usepackage{microtype} 
\usepackage{footmisc}  
\usepackage{algpseudocode}
\usepackage{xspace}
\usepackage{booktabs}
\usepackage{array}
\usepackage[ruled,linesnumbered]{algorithm2e}
\usepackage{wrapfig}
\usepackage{makecell}   
\usepackage{arydshln}   
\usepackage{diagbox}    
\usepackage{hyperref}
\usepackage{subcaption}

\usepackage{multirow}   
\usepackage{pifont}     

\usepackage{url}
\def\BibTeX{{\rm B\kern-.05em{\sc i\kern-.025em b}\kern-.08em
    T\kern-.1667em\lower.7ex\hbox{E}\kern-.125emX}}

\NewDocumentCommand{\var}{O{s} m O{}}{%
  \ensuremath{#1_{#2}^{#3}}
}
\usepackage{siunitx}
\usepackage{colortbl}

\newcommand{\commentout}[1]{}

\definecolor{light-gray}{gray}{0.80}

\newcommand\fref{Fig.~\ref}

\newcommand\sref{\S~\ref}

\usepackage{amsthm}

\usepackage{amsthm}

\newtheorem{mytheorem}{Theorem}
\newtheorem{assumption}[mytheorem]{Assumption}
\newtheorem{lemma}[mytheorem]{Lemma}
\newtheorem{remk}[mytheorem]{Remark}

\newcommand{\hide}[1]{}

\newcommand{\name}{MegaFold\xspace}

\newcommand{\nvidia}{NVIDIA\xspace}
\newcommand{\amd}{AMD\xspace}
\newcommand{\flashattention}{FlashAttention\xspace}
\newcommand{\torch}{PyTorch\xspace}

\newcommand{\bloom}{BLOOM\xspace}

\newcommand{\evo}{EvoAttention\xspace}
\newcommand{\swiglu}{SwiGLU\xspace}

\renewcommand{\emph}[1]{\textit{#1}}

\newcommand{\evoformer}{EvoFlash-3D\xspace}
\newcommand{\data}{EvoPipe\xspace}
\newcommand{\fusion}{EvoFusion\xspace}
\newcommand{\sdp}{EvoSP-3D\xspace}

\usepackage[skip=4pt]{caption}
\usepackage{setspace}
\usepackage{enumitem}

\setlist{nosep} 
\makeatletter 
\def\thm@space@setup{%
  \thm@preskip=3pt
  \thm@postskip=\thm@preskip 
}
\makeatother

\setlist[itemize]{noitemsep, topsep=0pt}

\makeatletter 
\renewenvironment{proof}[1][\proofname]{\par
  \vspace{1pt}
  \pushQED{\qed}%
  \normalfont
  \topsep0pt \partopsep0pt 
  \trivlist
  \item[\hskip\labelsep
        \itshape
    #1\@addpunct{.}]\ignorespaces
}{%
  \popQED\endtrivlist\@endpefalse
  \addvspace{3pt plus 3pt} 
}
\makeatother

\begin{document}

\title{MegaFold: Efficient Training of Next-Generation 3D Attention Protein Models on Cross-Platform GPUs}

\author{\IEEEauthorblockN{Hoa La\IEEEauthorrefmark{1}\IEEEauthorrefmark{2}}
\IEEEauthorblockA{\textit{University of Massachusetts, Amherst}\\
hvla@umass.edu}
\and
\IEEEauthorblockN{Ahan Gupta\footnote{Equal Contribution}\IEEEauthorrefmark{1}}
\IEEEauthorblockA{\textit{SSAIL Lab, University of Illinois Urbana-Champaign} \\
ag82@illinois.edu}
\and
\IEEEauthorblockN{Alex Morehead}
\IEEEauthorblockA{\textit{Lawrence Berkeley National Laboratory} \\
acmwhb@lbl.gov}
\and
\IEEEauthorblockN{Jianlin Cheng}
\IEEEauthorblockA{\textit{University of Missouri} \\
chengji@missouri.edu}
\and
\IEEEauthorblockN{Minjia Zhang}
\IEEEauthorblockA{\textit{SSAIL Lab, University of Illinois Urbana-Champaign} \\
minjiaz@illinois.edu}
\thanks{\IEEEauthorrefmark{1}Equal Contribution.}
\thanks{\IEEEauthorrefmark{2}Work done while interning at UIUC SSAIL Lab.}
}

\maketitle

\begin{abstract}
Recent advances in biomolecular modeling have been catalyzed by models such as AlphaFold3 (AF3), which introduce science-informed changes to the transformer architecture. Unlike transformers, a defining characteristic of AF3-style models is their 3D attention over 2D pairwise representations which produces tensors whose computation and memory costs scale \emph{cubically} with sequence length.  As a result, despite moderate parameter counts, AF3-style models are far more expensive to train than size-equivalent transformers, and are severely constrained by GPU memory capacity. Our characterization shows 3D attention fundamentally changes the training workload, causing massive 3D attention maps, complex inter-operator dependencies, kernel fragmentation, and heavy host-side data pipelines which differ substantially from LLM training, leading to poor utilization on modern GPU systems. Moreover, existing GPU optimizations do not adequately address these challenges due to complex cross-layer inter-operator dependencies introduced by 3D attention.
Motivated by these challenges, we introduce \name, a novel cross-platform system for efficient training of next-generation 3D-attention protein models. \name combines a memory-efficient 3D-attention kernel, a communication-efficient sharding strategy for quadratic representations, fused operator implementations for critical execution paths, and a determinism-aware host-device pipeline that eliminates preprocessing stalls. Evaluation on both \nvidia H200 and \amd MI250 GPUs shows that \name enables training with up to 3.36$\times$ longer sequence lengths on 32 GPUs while reducing end-to-end execution time by up to 1.73$\times$ (NVIDIA) and 1.62$\times$ (AMD). 

\end{abstract}

\begin{IEEEkeywords}
High performance computing, Bioinformatics, Parallel algorithms, Performance analysis.
\end{IEEEkeywords}

\section{Introduction}
\label{sec:introduction}

Deep learning has transformed structural biology, enabling accurate prediction of protein structures and biomolecular interactions~\cite{af2}. Models such as AlphaFold3 (AF3) extend this capability beyond single proteins to protein-protein, protein-DNA/RNA, and protein-ligand interactions~\cite{af3}, making a shift towards next-generation biomolecular modeling workloads~\cite{openfold,rosefold,esm-fold}. Unlike earlier models, these models incorporate science-informed architectural changes that fundamentally change the computational structure of training. 

A defining characteristic of AF3-style models is the use of 3D attention over 2D pairwise representations of amino acids. This design produces high-dimensional tensors whose compute and memory scale \emph{cubically} with sequence length. As a result, despite a moderate parameter count ($\sim$500M parameters)~\cite{af3}, AF3-style models are significantly more expensive to train than size-equivalent transformer models. In practice, AF3 training can be up to two orders of magnitude slower and supports input lengths that are an order of magnitude shorter. For example, sequences longer than 590 residues lead to out-of-memory (OOM) errors even with batch size one, while a size-equivalent 560M parameter transformer such as BLOOM-560M~\cite{scao2022bloom} easily scales to thousands of tokens on the same GPU.

To understand this gap, we perform a detailed performance characterization of AF3 training and identify four fundamental challenges rooted in 3D attention (\sref{sec:performance-characterization}).
\emph{Challenge-I:} AF3 training suffers from extreme memory pressure due to cubic scaling. Its core operator processes 2D pairwise representations and produces a 3D attention map with $O({N_{token}}^3)$ memory complexity which easily overwhelms on-chip caches and generates memory traffic that quickly exceeds device bandwidth limits. 
\emph{Challenge-II:} AF3 introduces multi-axis communication patterns that differ fundamentally from those in transformer training. To alleviate memory bottlenecks from long sequences, prior works develop sequence parallelism (SP) strategies for transformers~\cite{ulysses,megatron-SP}. However, extending these strategies to 2D pairwise representations requires sharding along both sequence and pair dimensions. This results in 3D communication flows that are not supported by conventional parallelism schemes. 
\emph{Challenge-III:} AF3 exhibits severe kernel fragmentation. To model complex biomolecular interactions, training interleaves 3D attention, transition layers, and projection operations, producing thousands of fine-grained kernels per iteration. These kernels are too small to efficiently utilize GPU execution pipelines, and lead to poor warp occupancy, underutilized memory bandwidth, and excessive intermediate activations.
\emph{Challenge-IV:} AF3 training suffers from heavy host-side preprocessing. Constructing the 2D pairwise inputs required by 3D attention involves multi-stage biological preprocessing, including multi-sequence alignment (MSA)~\cite{af2}, which runs on the CPU and introduces significant GPU idle time that cannot be hidden by conventional prefetching-based data pipelines.

Collectively, these challenges constrain AF3 training to short sequences and result in prohibitively long  training times (e.g., 11 days on 128 GPUs~\cite{cheng2023fastfoldreducingalphafoldtraining} despite being a moderate $\sim$500M parameter model). 
Although modern GPU systems provide highly optimized support for transformer training, these optimizations do not transfer to AF3. Techniques such as FlashAttention~\cite{flash-infer}, compiler-based fusion~\cite{chen2018tvm, ansor, pytorch-2}, and standard sequence parallelism~\cite{ulysses,ring-attention} assume 2D attention over 1D sequences and are insufficient for the cubic scaling, and cross-layer dependencies introduced by 3D attention.  

Motivated by these challenges, we present \name, a novel system for efficient training of next-generation 3D-attention protein models on modern GPUs. \name is the first end-to-end system we are aware of that provides integrated support for 3D attention training by jointly addressing kernel execution, parallelization, operator fusion, and host-device pipelining. \name introduces (1) \evoformer, a Triton-based kernel for 3D attention over 2D pairwise representations to achieve high memory and bandwidth efficiency on cross-platform GPUs (\sref{subsec:fused-attention}), (2) \sdp, a communication-efficient sharding strategy that supports alternating attention axes from 2D pairwise representations (\sref{subsec:sp-scheme}), (3) \fusion, a fused operator stack that aligns attention and transition layers with GPU shared-memory hierarchies (\sref{subsec:fused-small-ops}), (4) \data, a determinism-aware host-device pipeline that eliminates CPU-side preprocessing stalls (\sref{subsec:data-loading}). 

We implement \name on top of a widely used open-source AF3 training system~\cite{af3-open-source} and evaluate it on both \nvidia H200 and \amd MI250 GPUs. Our results show that \name reduces peak memory usage by up to 1.23$\times$, while improving per-iteration training time by up to 1.73$\times$ on \nvidia and 1.62$\times$ on \amd. Moreover, \name enables scalable training with up to 3.36$\times$ longer input sequence lengths on 32 GPUs. These results demonstrate that 3D-attention protein models introduce novel system-level challenges, and that dedicated system support is essential for scaling emerging biomolecular workloads on modern GPU platforms. 

\section{Background}
\label{sec:background}

\begin{figure}[t]
    \centering
    \includegraphics[width=0.93\linewidth]{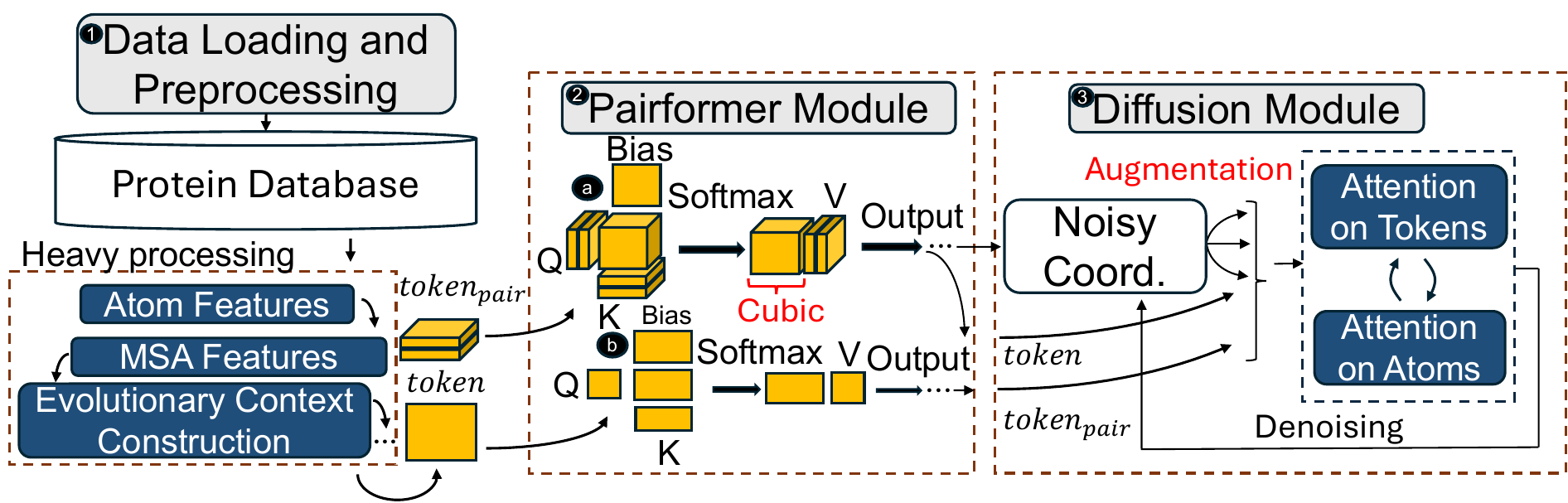}
    \caption{Overview of the AlphaFold3 (AF3). AF3 consists of 3 stages: (1) data loading and preprocessing, which constructs single-token and pairwise representations from input biomolecular complexes using evolutionary context; (2) the Pairformer backbone, which jointly processes 1D token and 2D pairwise representations using 3D attention with cubic complexity; and (3) a diffusion-based decoder, which generates 3D atomic coordinates through iterative denoising with data augmentation.} 
    \label{fig:af3-background} 
\end{figure}

In this section, we provide the necessary background on AlphaFold3 (AF3)~\cite{af3}, focusing on architectural features that give rise to its unique system challenges. 
AF3~\cite{af3} extends AF2~\cite{af2} from single-protein folding to modeling interactions among proteins, DNA, RNA, and small molecules. To support this generality, AF3 introduces a substantially more complex architecture and data pipeline than transformer-based language models. As shown in \fref{fig:af3-background}, AF3 consists of three stages: data loading and preprocessing, the Pairformer backbone, and a diffusion-based structure decoder. 

\noindent \textbf{Inputs and representations.}
AF3 processes raw biomolecular inputs into two primary internal representations. The \textit{single representation}, denoted as \texttt{token} with shape \texttt{[N$_{\text{token}},\text{h}_{\text{single}}$]}, which captures per-token features such as chemical identity and local geometry. To model interactions between tokens, AF3 also constructs a \textit{pair representation}, \texttt{token$_{\text{pair}}$} , with shape \texttt{[$\text{N}_{\text{token}},\text{N}_{\text{token}},\text{h}_{\text{pair}}$]}. This quadratic representation is initialized from geometric priors and enhanced using evolutionary context from multiple sequence alignment (MSA) and structural information from template complexes. These representations become the core inputs to the Pairformer module. 

\noindent \textbf{Pairformer and 3D attention.} The Pairformer module is one of the main computational backbones of AF3 and replaces Evoformer used in AF2.  It operates jointly on the 1D \texttt{token} representation and the 2D \texttt{token$_{\text{pair}}$} representation using two primary operators:  EvoAttention, and transition layers. EvoAttention is a 3D attention operator that computes attention over the 2D pair representation, while conditioning the computation on sequence-level features derived from the 1D token representation via broadcasted projections and bias terms. This mixed-dimensional attention produces intermediate tensors whose size scales cubically with sequence length, introducing substantial activation memory and complex data dependencies that differ from standard self-attention over 1D textual sequences.
The transition layers further process pairwise features using layer normalization and feed-forward networks with \swiglu~\cite{swiglu} activations, which preserves these mixed-dimensional dependencies across layers. 

\noindent \textbf{Diffusion-based structure decoding.} 
Following Pairformer, AF3 employs a diffusion-based decoder to generate 3D atomic coordinates conditioned on the learned representations. The decoder applies attention over atomic and token representations and iteratively denoises predicted structures. To improve robustness and learning signal, AF3 performs extensive data augmentation, which augments 3D atomic coordinates (e.g., by 48x) with random rotations and translations, increasing the effective batch size processed by the diffusion module.

\section{Performance Characterization}
\label{sec:performance-characterization}

We characterize four performance challenges by analyzing a widely used open-source AlphaFold3 implementation~\cite{af3-open-source}.

\subsection{3D Attention Induces Cubic Activation Memory Growth} 
\label{para:memory-heavy-layers} 
\begin{figure}[t]
    \centering
    \begin{minipage}{0.48\columnwidth}
        \centering
        \includegraphics[width=\textwidth]{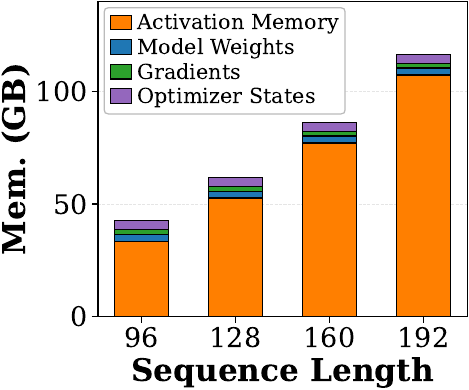}
        \caption{End-to-end memory breakdown of AF3 training across sequence lengths. Activation memory dominates memory consumption and grows rapidly with sequence length, while model parameters, gradients, and optimizer states remain relatively stable.}
        \label{fig:e2e_memory}
    \end{minipage}\hfill
    \begin{minipage}{0.48\columnwidth}
        \centering
        \includegraphics[width=\textwidth]{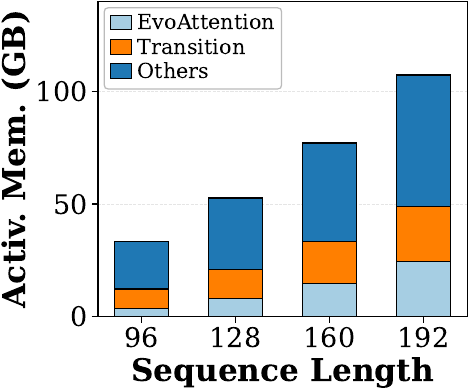}
        \caption{Breakdown of activation memory across operator groups in AF3. \evo exhibits fastest growth with sequence length due to its cubic tensors, while activations from transition and other operators increases and remains a substantial fraction of the total. }
        \label{fig:activation_memory}
    \end{minipage}
    \label{fig:combined_memory}
\end{figure}

AF3 has a moderate parameter count ($\sim$500M), yet incurs significant activation memory during training. This limits the maximum sequence length that can be supported and forces models to train relatively small biomolecular complexes. To quantify this effect, we measure AF3's end-to-end memory consumption across a range of sequence lengths in \fref{fig:e2e_memory}, 
breaking down memory consumption into model weights, activation, gradients, and optimizer states. 
We observe that while model weights, gradients, and optimizer states collectively consume only 4.08GB of memory, activation memory dominates overall usage, reaching up to 107.4GB, 96\% of total memory consumption. \fref{fig:activation_memory} further shows the breakdown of the activation memory by operator. 
Two operators, EvoAttention and Transition, are responsible for nearly $\sim$50\% of the total activation footprint, as they frequently appear across both the Pairformer and Diffusion modules. 
Among these, EvoAttention is the primary contributor to memory growth with increasing sequence length. At a sequence length of 96, EvoAttention produces $\sim$3.75GB of activations, which increases to $\sim$24.61GB at a sequence length of 192. This corresponds to a 6.56$\times$ increase in activation memory for a 2$\times$ increase in sequence length, reflecting the cubic complexity of 3D attention over 2D pairwise representations. As sequence length increases, the resulting activation tensors overwhelm on-chip caches and exceed GPU HBM capacity, leading to OOMs while also generating memory traffic that saturates device bandwidth even when execution is feasible. 

\smallskip
\noindent \fbox{\parbox{0.96\linewidth}{
\textbf{Insight-1:} 3D attention over 2D pairwise representations generates activations whose size grows cubically with sequence length. These activations dominate GPU memory usage, exceed HBM capacity at moderate sequence lengths, and stress cache locality and device bandwidth. }}
\smallskip

\subsection{2D Pairwise Representations Break Sequence Parallelism} 
\begin{figure}[t]
    \centering
    \begin{minipage}{0.48\columnwidth}
        \centering
        \includegraphics[width=\textwidth]{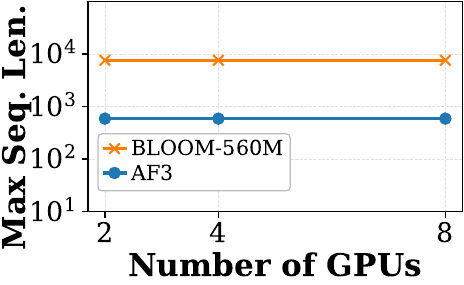}
        \caption{Max Seq. length trainable for AF3 and \bloom-560M under DDP as GPU count increases. Despite similar model sizes, AF3 supports substantially shorter sequences due to cubic activation in 3D attention.}
        \label{fig:seq-length-diff-hardware}
    \end{minipage}
    \hfill
    \begin{minipage}{0.48\columnwidth}
    \centering
    \includegraphics[width=\textwidth]{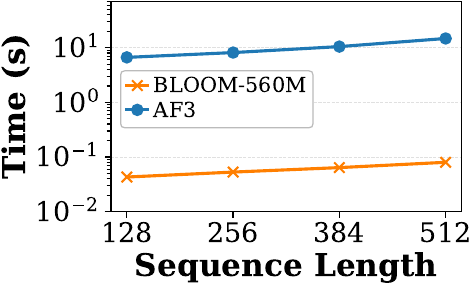}
    \caption{Average training iteration time of AF3 and \bloom-560M on a single GPU across sequence lengths. AF3 exhibits significantly higher per-iteration cost compared to the transformer-based \bloom-560M. }
    \label{fig:bloom-af3-step-time}
    \end{minipage}
    \label{fig:step-time-with-ddp}
\end{figure}

AF3 models are typically trained at scale with distributed data parallelism (DDP), largely because their moderate parameter count fits within the memory of a single GPU. While DDP effectively scales batch size, it does not scale input sequence length. We confirm this behavior by scaling AF3 training from 2 to 8  GPUs while holding the per-GPU micro-batch size fixed at 1 (\fref{fig:seq-length-diff-hardware}). Despite the additional GPU memory, training beyond a single sequence length of 590 residues consistently results in OOMs, indicating that conventional data-parallel scaling cannot translate increased hardware resources into longer sequence support for AF3. 
A natural next step would be to apply sequence parallelism (SP)~\cite{ring-attention, ulysses}, which has been shown to scale sequence length effectively for transformer models. However, extending SP to AF3 is substantially more complex due to its use of 2D pairwise representations \texttt{token$_\text{pair}$} (\texttt{[N$_\text{token}$,N$_\text{token}$,h$_\text{pair}$]}-sized) and attention operators that alternate their reduction axes across layers, which introduces communication and synchronization requirements that differ from those in 1D sequence models. As a result, supporting such an execution pattern requires sharding mechanisms that go beyond conventional SP designs developed for transformers. We describe such a design in \sref{subsec:sp-scheme}.  

\smallskip
\noindent \fbox{\parbox{0.96\linewidth}{
\textbf{Insight-2:} AF3 makes SP scaling substantially more complex than in transformers due to its use of 2D pairwise representations and attention operators with alternating reduction axes. As a result, extending SP to AF3 requires sharding and coordinating multi-dimensional tensors in ways that differs from the 1D SP designs commonly used for transformers.}}
\smallskip

\subsection{Mixed-Dimensional Operators Prevent Effective Kernel Fusion} 
\label{para:launch-overhead} 
\begin{figure}[t]
    \centering
    \includegraphics[width=\linewidth]{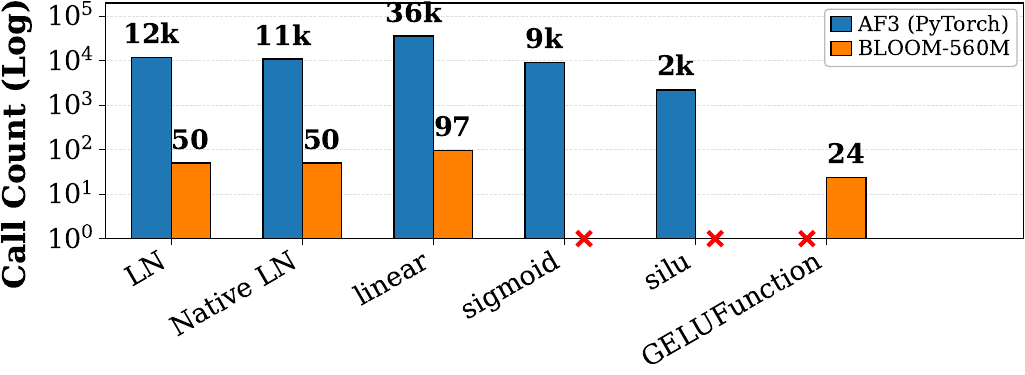}
    \caption{Kernel launch counts (log scale) for common operators in a single training iteration of AF3 and \bloom-560M. AF3 invokes orders of magnitude more small kernels than the size-equivalent transformer models. }  
    \label{fig:kernel-launch-characterization} 
\end{figure}

Despite AF3's moderate model size, training invokes a large number of fine-grained GPU kernels per iteration. This stems from the model's extensive use of mixed-dimensional operators that interleave computations over 1D sequence features, 2D pairwise representations, and higher-order intermediates. To quantify the resulting execution behavior, we profile kernel launches in a single AF3 training iteration and compare them against a size-equivalent transformer-based transformer model, \bloom-560M \cite{scao2022bloom}, as shown in \fref{fig:kernel-launch-characterization}.
We observe that AF3 launches approximately 36,000 linear-layer kernels, 23,000 layer-normalization kernels, and 11,000 activation-function kernels, over two orders of magnitude more than \bloom-560M. These kernels are small and often memory-bound, which reflect the fine-grained structure induced by repeatedly transitioning between operators defined over different tensor dimensionalities. As a result, AF3's execution graph is highly fragmented, preventing kernels from amortizing launch overhead or efficiently utilizing GPU execution pipelines. 

The fragmentation has direct consequences to both runtime and memory efficiency. Frequent kernel launches incur accumulated scheduling and dispatch overhead, while the inputs to \textit{each} kernel must be retained for gradient computation during the backward pass, increasing activation memory pressure. \fref{fig:bloom-af3-step-time} shows the average per-iteration time of AF3 and \bloom-560M across different sequence lengths. Across all configurations, AF3's iteration time is consistently two orders of magnitude higher than that of the size-equivalent transformer.
The root cause of this behavior lies in the structure of AF3's operators. AF3 interleaves 3D attention, transition layers, triangular updates, and normalization operations that operate over different dimensional representations and exhibit non-trivial reduction semantics. As a result, reducing this fragmentation, whether through fusion or other optimizations, needs to coordinate reductions, broadcasts, and data layouts across operators with incompatible dimensional assumptions.

\smallskip
\noindent \fbox{\parbox{0.96\linewidth}{
\textbf{Insight-3:} AF3's extreme kernel fragmentation is a structural consequence of its mixed-dimensional execution. Training interleaves operators over 1D sequence features, 2D pairwise representations, and higher-order intermediates, creating fine-grained kernels with incompatible reduction and data layout semantics. These cross-operator dependencies prevent kernels from being easily fused together, introducing significant kernel launch and activation overhead.}}
\smallskip

\subsection{Multi-Representation Evolutionary Context Construction Slows the Data Pipeline} 
\label{para:data-loading} 

\begin{figure}[t]
    \centering
    \includegraphics[width=\linewidth]{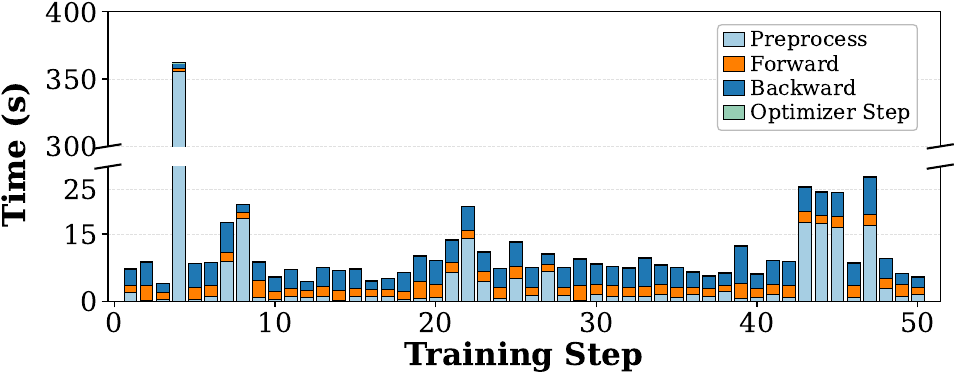}
    \caption{Breakdown of AF3 training iterations into data-loading and preprocessing, forward pass, backward pass, and optimizer step. Data loading and preprocessing exhibits variable overhead across iterations, leading to frequent GPU idle time.}
    \label{fig:data-loading-idling}
\end{figure}

Although AF3 operates on relatively short biomolecular complex lengths due to its large activation memory footprint, host-side data loading and preprocessing consume a substantial fraction of end-to-end training time. This overhead primarily comes from the construction of evolutionary context required to generate 2D pairwise representations used by EvoAttention. 
We profile 50 iterations of an AF3 training loop and decompose each iteration into data-loading and preprocessing (DLP), forward pass, backward pass, and optimizer step, as shown in \fref{fig:data-loading-idling}. On average, data-loading and preprocessing consumes 38.2\% of total iteration time, excluding a single outlier caused by unusually long MSA ($\sim$16k residues). Moreover, there is a large variance in runtime across iterations due to differences in input sequence lengths and evolutionary context, leading to frequent GPU idle periods. 

The GPU idling stems from the complex dependencies and I/O-bound preprocessing. AF3's DLP stage prepares multiple representations: pairwise representations for EvoAttention processing, alongside single, MSA, template, atomic and tokenized representations for processing by other operators. Certain representations have a dependency on the creation of another (e.g. tokenized representations are aggregated through atomic representations) with I/O-bound retrieval to gather similar templates of the input complex (to enrich evolutionary contexts) dominating runtime. These sequential dependencies and I/O-bound retrieval steps make applying data-prefetching optimizations to mitigate host-side delays challenging. Increasing the number of data loading process only marginally reduces the per-iteration time. 

\smallskip
\noindent \fbox{\parbox{0.96\linewidth}{
\textbf{Insight-4:} Constructing evolutionary context for multiple representations introduces dependency-heavy and IO-bound preprocessing, causing persistent GPU idle time that cannot be easily eliminated by conventional data-prefetching. 
}}
\smallskip

\section{MegaFold Design}
\label{sec:megafold-design}

\begin{figure*}[!ht]
    \centering
    \includegraphics[width=1\linewidth]{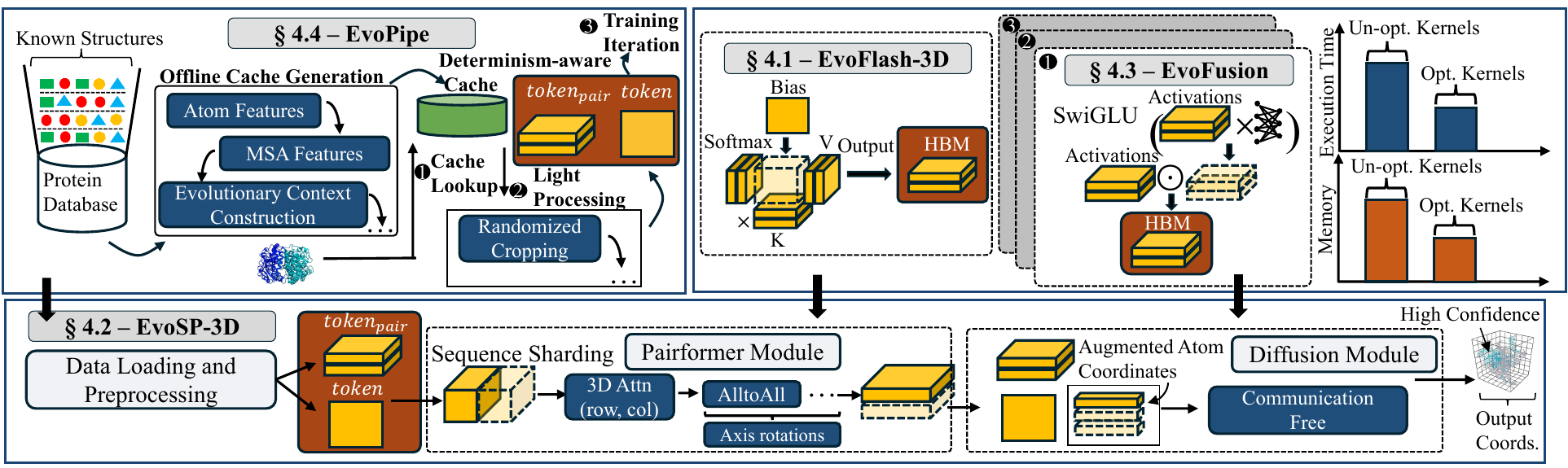}
    \caption{Overview of \name and how its optimizations are coordinated across the training pipeline. \data leverages determinism-aware cache mechanism to produce \texttt{token} and \texttt{token$_{\text{pair}}$} representations. These representations flow through the Pairformer and diffusion stages, where \evoformer executes memory-efficient 3D attention,  \sdp shards pairwise representations and coordinates inter-device communication, and \fusion reduces kernel fragmentation in transition layers across stages. Together, these optimizations enable efficient long-sequence AF3 training. 
    }
    \label{fig:megafold-overview}
\end{figure*}

To address the bottlenecks identified in \sref{sec:performance-characterization}, \name adopts a design that directly targets the sources of inefficiency in AF3 training. \fref{fig:megafold-overview} shows an overview of \name: (1) \evoformer addresses the cubic activation memory and bandwidth pressure introduced by 3D attention~\sref{subsec:fused-attention}, (2) \sdp enables longer sequence training over 2D pairwise representations by restructuring parallelism~\sref{subsec:sp-scheme}, (3) \fusion reduces kernel fragmentation caused by mixed-dimensional operators~\sref{subsec:fused-small-ops}, and (4) \data mitigates host-side stalls introduced by evolutionary context construction~\sref{subsec:data-loading}. Together, these components form a cross-stack system that enables efficient AF3 training on modern GPU platforms. 

\subsection{\evoformer: Cross-Platform Implementation of Memory-Efficient 3D Attention}
\label{subsec:fused-attention}

AF3's EvoAttention operator is the dominant source of activation memory pressure during training. As shown in \sref{para:memory-heavy-layers}, EvoAttention operates on 2D pairwise representations and produces 3D attention map that scales cubically with sequence length. Supporting long-sequence training therefore requires a kernel design that avoids materializing these intermediate tensors while preserving the correctness of EvoAttention in both forward and backward gradient computation. 

Designing such a memory-efficient kernel is non-trivial. EvoAttention normalizes attention logits using softmax before computing the output, which conventionally requires materializing the full attention matrix. Prior work such as \flashattention~\cite{dao2022flashattention}, addresses this issue by algorithmically restructuring attention calculation using online, numerically stable softmax~\cite{online-softmax}. However, EvoAttention differs fundamentally in that it operates over 2D pairwise inputs and introduces an additive broadcasted pair-bias term. This results in additional broadcast operations in the forward pass and extra reduction operations in the backward pass, introducing multi-dimensional data dependencies that requires a specialized kernel design.

We introduce \evoformer, a memory-efficient kernel for EvoAttention that avoids materializing the cubic attention logits in global memory. Instead, \evoformer incrementally computes attention logits within fast scratchpad memory during both forward and backward passes. This design asymptotically reduces the peak activation memory of EvoAttention from \texttt{O(N$_{\text{token}}^3$)} to \texttt{O(N$_{\text{token}}^2$)} while maintaining numerical correctness. \evoformer is implemented using Triton~\cite{triton} and is therefore performance-portable.

\begin{figure}[t]
    \centering
    \includegraphics[width=\linewidth]{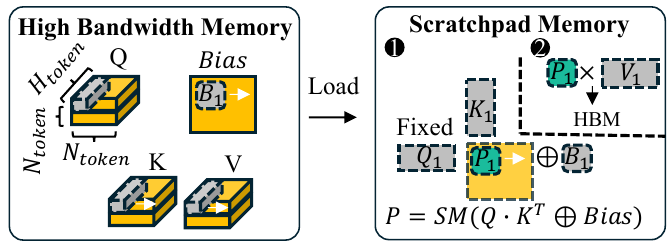}
    \caption{\evoformer's forward pass on a 2D pairwise input. Green boxes represent computations performed within a  thread block using scratchpad memory, while gray boxes indicate accesses to queries, keys, values \& bias stored in HBM. The attention-map, $P$, is accumulated on the fly and never materialized in HBM, reducing peak memory of \evo.}
    \label{fig:fused-attn}
\end{figure}

\fref{fig:fused-attn} illustrates the forward pass of the kernel, demonstrating a thread-block's access patterns over the \verb|Q|, \verb|K| \& \verb|V| tensors as well as the rows of the attention logits a block computes.
It launches 2D thread blocks: \texttt{[b$_y$, b$_x$]}, with each block computing one ``chunk" (consisting of multiple rows) of the output tensor $o$. Because EvoAttention requires normalizing attention logits prior to being multiplied by the \texttt{values}, \evoformer adopts an online softmax strategy~\cite{online-softmax}.
Listing~\ref{alg:fused-attn} shows the forward pass as \verb|fused_attn_fwd|. Each thread block maintains lightweight metadata, which consists of a running normalization factor \texttt{li} and a running maximum \texttt{mi}, both initialized locally as \texttt{0} and \texttt{-inf} respectively (line 9). During execution, the block first loads its assigned query vectors $Q$ (line 10), then iterates over tiles of the key and value tensors. In each iteration, the kernel computes the query-key products and applies the corresponding pairwise bias via on-the-fly broadcasting (line 13-14). The attention logits are shifted by the current running maximum  (line 16), after which the metadata (\verb|li|, \verb|mi|) are updated to maintain numerical stability (line 20). Partial outputs \verb|oi| are normalized using the \texttt{mi} - running sum - metadata  (line 21-22), without ever materializing the full attention matrix. After all key-value tiles are processed, the accumulated outputs are normalized using the final metadata values. The resulting log-sum-exp terms are stored to enable correct re-normalization during the backward pass. 

The backward computation is decomposed into two stages to accommodate different parallel strategies required to compute the gradient tensors. One stage computes the gradients with respect to the value and key tensors (\texttt{dV} \& \texttt{dK}), while the other computes gradients for the queries and pairwise bias terms (\texttt{dQ} \& \texttt{dBias}). Notably, \texttt{dBias} gradients require reductions across attention dimensions. This is achieved by repeatedly accumulating local partial gradients into the corresponding global location via \texttt{atomic\_add}s. This strategy avoids retaining partially reduced \texttt{dBias} gradients in scratchpad memory and preserves kernel occupancy. 


\definecolor{codered}{rgb}{0.6,0,0}
\definecolor{codegreen}{rgb}{0,0.6,0}
\definecolor{codeblue}{rgb}{0,0,0.6}
\definecolor{codepurple}{rgb}{0.58,0,0.82}
\definecolor{codebg}{rgb}{0.97,0.97,0.97}

\lstdefinestyle{fused-attn-style}{
    language=Python,
    frame=single,
    framerule=0pt,
    backgroundcolor=\color{codebg},   
    commentstyle=\color{codepurple},
    keywordstyle=\color{codeblue},
    emphstyle=\ttfamily\color{codered},
    stringstyle=\color{codegreen},
    basicstyle=\ttfamily\linespread{0.85}\footnotesize,
    breakatwhitespace=false,         
    breaklines=true,                 
    captionpos=b,                    
    keepspaces=true,                 
    numbers=left,                    
    numbersep=-4pt,                  
    showspaces=false,                
    showstringspaces=false,
    showtabs=false,                  
    tabsize=2,
    morekeywords={}
}

\lstset{
  mathescape,         
  literate={->}{$\rightarrow$}{2}
           {ε}{$\varepsilon$}{1}
}

\begin{lstlisting}[style={fused-attn-style}, 
    label={alg:fused-attn},
    caption={Pseudocode for the memory-efficient fused EvoAttention kernel using EvoFlash-3D. The kernel computes 3D attention over 2D pairwise inputs using stable online softmax, accumulating attention outputs on the fly and avoiding materialization of the cubic attention tensor in global memory.},
    float=t]
  # Omit indexing on batch/head/hid. dim for brevity
  def fused_attn_fwd(Q,K,V,mask,bias):
    """Launched with thread-blocks of size: [by, bx]
    Q, K, V: [b, heads, N, N, d] tensor.
    mask: [b, N, N] tensor.
    bias: [b, heads, N, N] tensor.
    """
    i = tl.program_id(0)
    li, mi = [0]*by, [-inf]*by
    qi = Q[i*by:(i+1)*by, :]
    oi = zeros((by, d))
    for j in range(N/bx): 
      s = tl.matmul(qi, k[j*bx:(j+1)bx,:].T)+\   
            bias[i*by:(i+1)by,j*bx:(j+1)bx]
      mij = tl.maximum(mi,tl.max(s,1))
      s -= mij
      p = tl.exp(s)
      lij = tl.sum(s, axis=-1)
      alpha = tl.exp(mi-mij)
      li, mi = li*alpha+lij, mij
      oi = oi * alpha + \
            tl.matmul(p,v[j*bx:(j+1)bx, :])
    return oi, li, mi\end{lstlisting}

\subsection{\sdp: Communication-Efficient Sharding For 2D Pairwise Representations}
\label{subsec:sp-scheme}

As shown in \sref{sec:performance-characterization}, SP is not directly applicable to AF3 training due to the use of mixed-dimensional representations and inter-operator dependencies. We therefore design \sdp, a parallelization strategy tailored to 2D pairwise representations, which enables scalable training without repeated resharding or synchronization between operators. 

\begin{figure}[t]
    \centering
    \includegraphics[width=\linewidth]{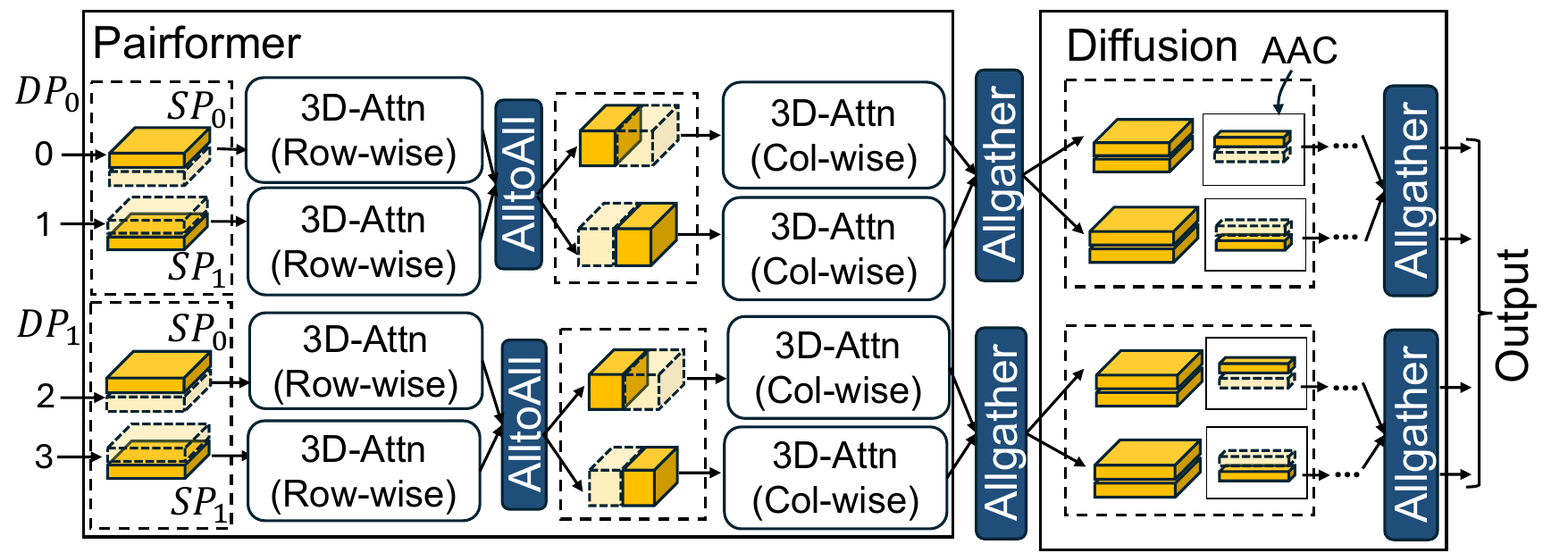}
    \caption{\sdp with two DP and SP groups. Within Pairformer, successive \evo operate on different axes of the 2D  \texttt{token$_\text{pair}$} representation. \sdp uses \texttt{all\_to\_all} to rotate the sharding layout at 3D attention boundaries so that each 3D attention receives the required 2D slices. In the diffusion stage, augmented atom coordinates (AAC) are distributed across SP groups without communication.}
    \label{fig:dp-sp-hybrid}
\end{figure}

In \sdp, DP groups operate on different inputs within the batch, while the notion of SP groups is reused as a coordination mechanism to partition and rearrange 2D pairwise representations across stages of AF3's execution pipeline. Multiple SP groups form a DP group, enabling batch-size scaling while allowing stage-specific partitioning tailored to AF3's mixed-dimensional operators. 
We illustrate the execution of \sdp using a 2x2 (DPxSP) configuration in \fref{fig:dp-sp-hybrid}. During data loading and preprocessing, two tensors are produced, a sequence-level representation \texttt{token} with shape \texttt{[N$_{\text{token}}$,h$_{\text{single}}$]}, and a pairwise representation \texttt{token$_{\text{pair}}$} with shape \texttt{[N$_{\text{token}}$,N$_{\text{token}}$,h$_{\text{single}}$]}. Because \texttt{token} is relatively small, it is replicated within each DP group prior to Pairformer execution. In contrast,  \texttt{token$_{\text{pair}}$} is initially partitioned along one sequence axis, with different SP groups operating on disjoint contiguous shards. This is communication-free, as each group drops a different portion of the sequence.

Within the Pairformer module, execution is organized as a sequence of pipeline stages operating on the 2D \texttt{token$_{\text{pair}}$} representations. Each stage applies \evo with a fixed sharding axis of the \texttt{token$_{\text{pair}}$} tensor. Across successive stages, \evo alternates the semantic axis along which attention is computed, requiring the sharding axis to rotate accordingly. To transition between stages with different sharding axes, \texttt{all\_to\_all} collectives are used to rearrange the layout of \texttt{token$_{\text{pair}}$}. Each \texttt{all\_to\_all} exchanges data proportional to the locally held shard, resulting in an aggregate communication volume proportional to \texttt{O(N$_{\text{token}}^\text{2}$h$_{\text{token}}$)}. These collectives transform the layout of \texttt{token$_{\text{pair}}$} from one stage into the layout required by the next, enabling successive \evo stages to operate on the correct 2D slices without materializing the full quadratic tensor. 
In addition, \evo's additive bias term, derived from \texttt{token$_{\text{pair}}$}, is replicated across device via an \texttt{all\_gather} that is launched asynchronously and overlapped with computation, avoiding additional critical path overhead. 
After Pairformer processing, the updated \texttt{token$_{\text{pair}}$} tensor is passed to the diffusion module together with a noised atom-coordinate tensor. 

After Pairformer processing, the updated \texttt{token$_{\text{pair}}$} tensor is passed to the diffusion module together with augmented atom coordinates. As described in \sref{sec:background}, diffusion module applies random rotations and translations to generate multiple noised atom coordinates per input. 
\sdp exploits this structure by distributing augmented atom coordinates across SP groups, while \texttt{token$_{\text{pair}}$} is \texttt{all\_gather}ed so that diffusion operates on the full pairwise representation. During the forward pass, each SP group processes its local augmented atom coordinates independently without communication. After diffusion, denoised atom coordinates are \texttt{all\_gathered} on all SP groups to reconstruct the full output. 

By allowing partitioning and communication patterns to vary across stages, \sdp avoids repeated reshaping and synchronization between incompatible representations, which would otherwise arise from naively applying SP. This design enables scalable execution over 2D pairwise representations while preserving batch-level scaling through DP, making it well suited to AF3's training execution patterns.

\subsection{\fusion: Fused Operator Stack}
\label{subsec:fused-small-ops}

As shown in \sref{para:launch-overhead}, AF3 training suffers from kernel fragmentation, particularly in transition layers that interleave normalization, linear projection, and activation operators.
These operators are individually lightweight, but are executed as thousands of fine-grained kernels per iteration, leading to substantial kernel launch overhead, intermediate activation materialization, and excessive memory traffic. 

A key challenge is that AF3's transition layers operate  over mixed-dimensional representations and involve distinct computation patterns. Specially, layer normalization performs reductions over the hidden dimension, linear layers perform dense matrix multiplication over pairwise tensors, and activation functions apply element-wise transformations. These operators therefore have different iteration spaces, data access patterns, and synchronization requirements. Although epilogue/prologue fusion~\cite{pytorch-2} and schedule-rewrite based techniques~\cite{chen2018tvm} can compose such operators at their boundaries, doing so forces them into a single execution schedule that is ill-suited for at least one component. In practice, this often degrades performance by increasing register and shared-memory pressure, reducing occupancy, or introducing inefficient reduction strategies, offsetting the benefits or reduced kernel launches. 
Consequently, addressing kernel fragmentation in AF3 requires fusion strategies that go beyond composing operators at their boundaries. Effective optimization must reorganize operator internals to preserve efficient reduction and GeMM schedules while eliminating redundant intermediate materialization. 

To address these issues, we introduce \fusion, a fused operator stack that targets the most frequent and performance-critical operators in AF3. \fusion provides fused Triton kernels for layer normalizations (layer-norms), linear projections, and \swiglu activations, which collectively account for a large fraction ($\sim$20\%) of activation memory and kernel invocation during training. Rather than composing operators at their boundaries, \evoformer restructures their internal execution to reduce kernel launches and intermediate tensor materialization while preserving efficient computation schedules. 

\begin{figure}[t]
    \centering
    \includegraphics[width=\linewidth]{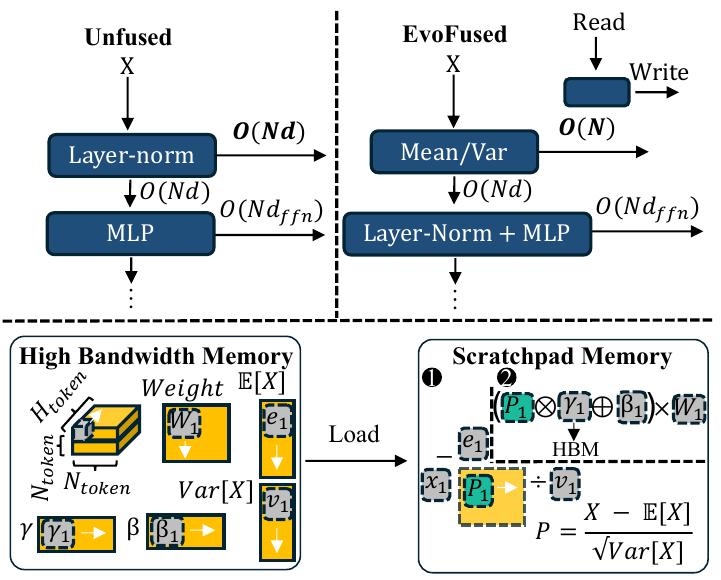}
    \caption{\fusion's forward pass for layer-norm \& linear-layer fusion compared to unfused counterpart. Gray and green blocks represent values read and computed by a thread-block, respectively. By modifying inner-matmul loops, \fusion reduces the data written to memory.} 
    \label{fig:lin-layer-fusion}
\end{figure}

\noindent \textbf{Fused layer-norm \& linear projection.} 
The layer-norm and linear project fusion implementation follows a two-stage design, as shown in \fref{fig:lin-layer-fusion}. In the forward pass, the first kernel computes per-row mean and variance of the input \verb|token| tensor. A thread-block operates on an entire row in \verb|token|, with $M$ thread-blocks launched ($M$ being the number of rows in \verb|token|).
A second kernel then performs normalization and matrix multiplication to compute the final output, 
\texttt{o$_{\texttt{2}}$}, with each thread-block computing a single \texttt{[b$_{\texttt{m}}$, b$_{\texttt{n}}$]}-sized tile of \texttt{o$_{\texttt{2}}$}. Each thread-block proceeds by first zero-initializing a \texttt{[b$_{\texttt{m}}$, b$_{\texttt{n}}$]}-sized tile, where its local answer of \texttt{o$_{\texttt{2}}$} will be computed. Then, an inner-loop over the hidden dimension first loads a chunk of the input \texttt{token} tensor - \texttt{c$_\text{t}$} - layer-norm data (mean, variance of input \& $\gamma$, $\beta$ parameters) - \texttt{c$_\mathbb{E}$}, \texttt{c$_\text{Var}$} \& \texttt{c$_\gamma$}, \texttt{c$_\beta$} - and linear-layer weights - \texttt{c$_\text{w}$}. Using these values to first normalize the input chunk, computing: ((\texttt{c$_{\text{t}}$} - \texttt{c$_\mathbb{E}$})/\texttt{c$_\text{Var}$}) $\otimes$ \texttt{c$_\gamma$} + \texttt{c$_\beta$}, and then finally multiplying these normalized values with the linear-layer weights. Structuring the forward pass in this manner, as opposed to computing everything in a single kernel, reduces shared-memory pressure as fewer intermediate values need to be stored, and results in good occupancy. Moreover, it avoids storing a \texttt{token$_\text{pair}$} sized tensor to global memory, achieving the best of both worlds: reducing memory while increasing speed. 
The backward pass follows a similar structure.

\noindent \textbf{Fused \swiglu.} 
The \swiglu activation applies an element-wise gating function of \texttt{token$_{\text{pair}}\oplus$sigmoid(token$_{\text{pair}}$) $\oplus$X}, where \texttt{X = token$_{\text{pair}}$W$_\text{1}$} and \texttt{W$_\text{1}$} are the weights of a linear-layer. 
\fusion implements \swiglu as a single fused kernel for the forward pass, and a single kernel for the backward pass, eliminating intermediate activations and reducing kernel launch overhead while preserving per-row parallelism.

\noindent \textbf{Fused transition layer.} Together, the fused layer-norm-linear kernel, the fused \swiglu activation, and the subsequent linear projection form a restructured transition layer. By eliminating materialization boundaries across these operators, \evoformer reduces kernel fragmentation and activation memory while preserving efficient execution schedules for training.

\subsection{\data: Determinism-Aware Host-Device Pipeline}
\label{subsec:data-loading}

As shown in \sref{para:data-loading}, AF3 training incurs substantial host-side overhead due to dependency-heavy and IO-bound preprocessing for constructing evolutionary contexts, leading to frequent GPU idle time. This overhead arises from science-specific preprocessing steps such as RDKit-based atom construction, the generation of MSA, and template processing, which are executed on the CPU and sit on the critical path of training. \data is designed to mitigate this bottleneck by decoupling deterministic preprocessing from the per-iteration training loop.

An AF3 training iteration consists of three stages. In the first stage, the input biomolecular complex is analyzed to construct atomic-level representations, such as atom shapes and coordinates, and to parse and tokenize MSA files to extract evolutionary features. These operations rely on expensive RDKit routines and often process long MSAs, making this stage both compute-intensive and highly variable in runtime. In the second stage, atomic and MSA features undergo additional preprocessing, such as random cropping to a target sequence length and tokenization. Finally, in the third stage, the resulting tokenized representations are consumed by the model to perform a forward and backward pass. 

\data exploits the observation that a subset of operations in the first stage produce deterministic outputs for a given input complex. In particular, RDKit-based atom construction and the conversion of raw MSA files into feature tensors do not depend on stochastic augmentation. \data precomputes these deterministic features once prior to training and stores them in a cache that maps each input complex to its corresponding atomic and MSA representations. During training, these cached features are retrieved through lightweight lookups and aligned with the current input sequence, which eliminates repeated execution of expensive preprocessing steps. 

Non-deterministic operations, such as random cropping and data augmentation, remain in the per-iteration training loop and are applied on top of the cached features. By separating deterministic preprocessing from iteration-time execution, \data significantly reduces host-side overhead and variability, enabling more consistent GPU utilization during AF3 training.

\section{Evaluation}
\label{sec:evaluation}

We evaluate \name to answer the following questions: (1) does it significantly improve the trainability and performance of AF3-style models, (2) does it scale efficiently across multiple GPUs, and (3) how much does each system optimization contribute to the end-to-end gains. We report end-to-end training performance on both NVIDIA and AMD GPUs.

\subsection{Evaluation Methodology}

\noindent \textbf{Hardware Testbed.} We evaluate \name on two GPU platforms: \nvidia H200-141GB and \amd MI250-64GB GPUs. Each NVIDIA node contains 8$\times$H200 GPUs connected via NVLink, while each AMD node contains 4$\times$ MI250 GPUs. 

\noindent \textbf{Datasets.} We use the same data sources and preprocessing pipeline as AF3-style models, including experimentally determined structures from Protein Data Bank (PDB)~\cite{bank1971protein}, together with MSA and template information required for \evo. The dataset covers a diverse set of protein and biomolecular complexes with varying sequence lengths. 

\noindent \textbf{Implementation.} We implement \name on top of a widely used open-source framework AlphaFold3-PyTorch~\cite{af3-open-source}. We use \torch 2.7.0, Triton 3.2.0 with CUDA 11.8 on \nvidia and ROCm 6.3 on \amd GPUs. \data has a bounded host-side cache for evolutionary context construction. In our experiments, this cache occupies 395GB disk space for the PDB dataset and is built offline in approximately 48 hours using 16 CPU for the entire AF3 training dataset (127,234 input complexes). Cache construction is embarrassingly parallel across input complexes and can be significantly accelerated by using additional CPUs. Moreover, this preprocessing step is amortized across training iterations and does not lie on the critical path of GPU execution. We observe no measurable impact on end-to-end scalability from \data cache management. 

\noindent \textbf{Baselines.} We compare MegaFold against the default AlphaFold3-PyTorch training pipeline~\cite{af3-open-source} executed in PyTorch Eager Mode, which represents the reference implementation used in practice. On NVIDIA GPUs, we additionally compare against DeepSpeed-Evoformer~\cite{song2023deepspeed4science}, which provides hand-optimized CUTLASS kernels specialized for AlphaFold-style workloads. For completeness, we also evaluate \torch Inductor~\cite{pytorch-2} as a representative compiler baseline.

\noindent \textbf{Training Configuration.} Unless otherwise stated, all experiments use a batch size of 1 with activation checkpointing (AC)~\cite{chen2016trainingdeepnetssublinear} for both \name and baselines. This configuration follows prior AF3 training practice and is needed to support practical sequence lengths under GPU memory constraints. We conduct all experiments on AF3 using the same optimizer and hyperparameters as described in AF3~\cite{af3}. For multi-GPU experiments, we additionally use distributed data parallelism (DDP)~\cite{ddp}.
For \name, we set \sdp with a DP group size of 2, and SP group sizes that scale with GPU count (e.g., 2, 4, 8, 16, 32) unless otherwise noted. Baseline systems use pure DP with the DP group size equal to the total number of GPUs.

\noindent \textbf{Metrics.} We evaluate \name using metrics that capture end-to-end performance, memory efficiency, and scalability. These include per-iteration training time, TFLOPS, peak GPU memory usage, strong/weak scaling efficiency and Model Flops Utilization (MFU). MFU measures the efficiency of an AI training workload, quantitatively measuring the fraction of throughput a workload achieves over the hardware's peak, accounting for only useful work (removing FLOPs due to AC).

\subsection{Main Results}
\label{subsec:main-results}

We evaluate \name's end-to-end training performance on AF3-style models, focusing on three key metrics: (1) maximum trainable sequence length, which determines the size of biomolecular complexes that can be modeled during training, (2) per-iteration execution time, and (3) peak GPU memory consumption. We compare \name against PyTorch Eager (PyTorch) and PyTorch Inductor (Inductor) baselines on both NVIDIA H200 and AMD MI250 GPUs. 

\subsubsection{Maximum Trainable Sequence Length} 
\begin{figure}[t]
    \centering
    \begin{minipage}{0.48\columnwidth}
        \centering
        \includegraphics[width=\textwidth]{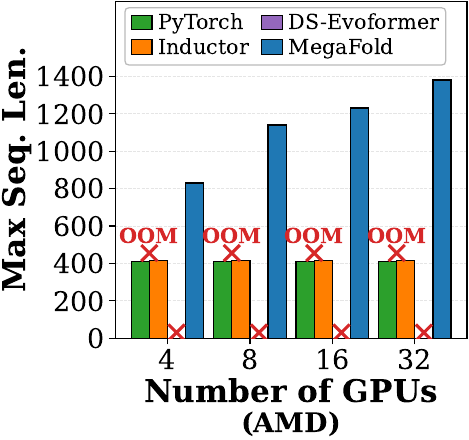}
        \caption{Maximum trainable sequence length before OOMs for different systems as GPU count increases. \name scales to substantially longer sequences than PyTorch baselines, while DS-Evoformer is unsupported on AMD.}
        \label{fig:seq-length-scaling}
    \end{minipage}\hfill
    \begin{minipage}{0.48\columnwidth}
        \centering
        \includegraphics[width=\textwidth]{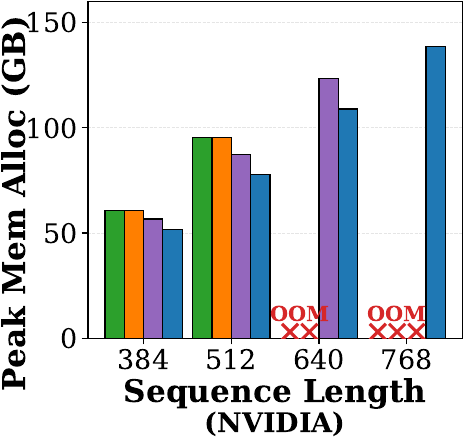}
        \caption{Per-device peak memory consumption prior to OOMs across sequence lengths on NVIDIA GPUs. \name reduces peak memory usage compared to PyTorch baselines, enabling training at longer sequence lengths. }
        \label{fig:memory-usage}
    \end{minipage}
    \label{fig:trainibility-results}
\end{figure}

\fref{fig:seq-length-scaling} shows the maximum trainable sequence length under different GPU counts. \name substantially improves max sequence trainability, enabling up to 3.36$\times$ longer input sequences compared to PyTorch baselines on 32 AMD GPUs. While baseline systems encounter OOM failures at moderate sequence lengths due to cubic activation growth in 3D attention, \name consistently supports longer sequences. DeepSpeed-Evoformer (DS-Evoformer)  relies on hand-written CUTLASS kernels and is therefore only evaluated on NVIDIA GPUs, as it is not compatible with AMD platforms.
The gains in maximum trainable sequence length come from two complementary effects. On a single GPU, \evoformer and \fusion reduce the activation footprint, enabling up to 35\% longer sequence (\fref{fig:optimization-ablation}, left). In multi-GPU settings, \sdp further extends this boundary by sharding quadratic pairwise representations across devices, enabling another up to 3$\times$ longer sequence. Together, kernel-level memory optimizations and distributed sharding substantially shift the OOM barrier. 

\begin{figure}[t]
    \centering
    \includegraphics[width=\linewidth]{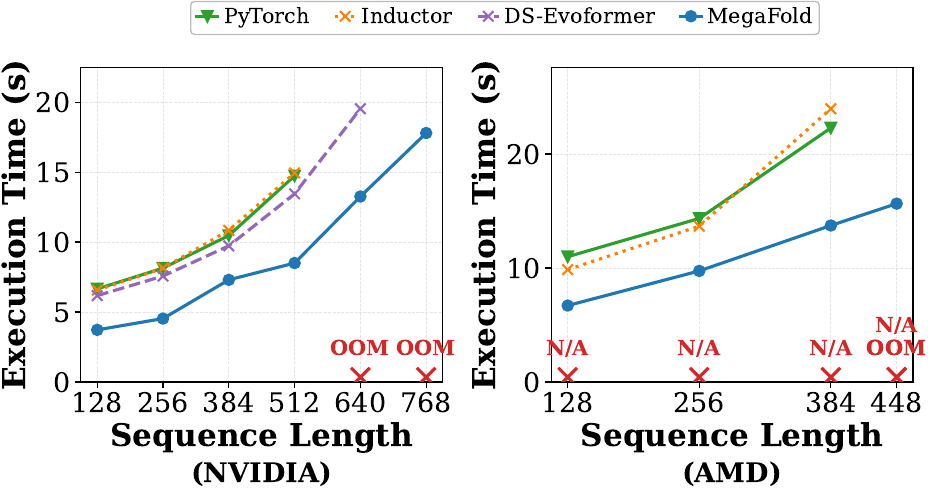}
    \caption{Average per-iteration execution time of AF3 training on a single GPU across sequence lengths. Results are shown on NVIDIA (left) and AMD (right) hardware. \name achieves speedups and supports longer sequences compared to baselines.}
    \label{fig:speed-results}
\end{figure}

\subsubsection{End-to-End Speed} \fref{fig:speed-results} reports the comparison of per-iteration training time across a range of sequence lengths. \name reduces iteration time by up to 1.73$\times$ on NVIDIA H200 and 1.62$\times$ on AMD MI250 compared to PyTorch Eager, while simultaneously supporting longer sequences.
In contrast, PyTorch Inductor provides little benefit for AF3 workloads due to graph breaks introduced by mixed-dimensional operators, which limit cross-operator fusion. 
Performance gains grow with sequence length, which reflects the increasing effectiveness of \name's memory-efficient kernels and fused operators under high activation pressure. 

\subsubsection{Memory Efficiency} \fref{fig:memory-usage} shows peak GPU memory usage during training. \name reduces peak memory consumption by up to 23\% relative to the PyTorch baselines, allowing memory savings to be reallocated toward longer sequences. As a result, \name is able to train up to sequence length 768, whereas baselines run OOMs. These reductions stem from avoiding materialization of cubic intermediate tensors in 3D attention and eliminating redundant intermediate activations in transition layers. 

The evaluation shows that \name's performance gains align with its system design. Memory-efficient 3D attention \evoformer and fused transition operators via \fusion shift the OOM barrier to support longer sequences, while \sdp and \data improve training scalability and speed. Together, these techniques enable efficient, cross-platform training of AF3-style models.

\subsection{Scalability Evaluation}
\label{subsec:multi-gpu}

We evaluate \name's scalability to assess whether its communication and pipeline design remains efficient as GPU count increases. We report both strong and weak scaling results on AMD MI250 GPUs, comparing \name against PyTorch Eager Mode, which provides the strongest baseline on this platform. All experiments use a fixed sequence length of 384 with \sdp set to 2.

\begin{figure}[t]
    \centering
    \begin{minipage}{0.48\columnwidth}
        \centering
        \includegraphics[width=\textwidth]{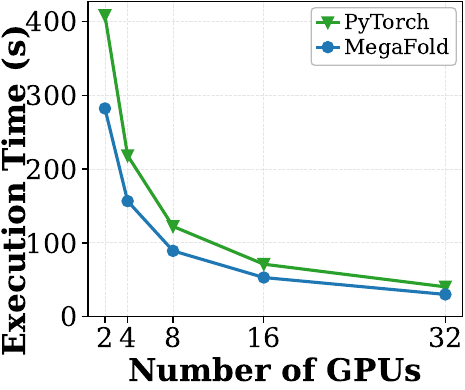}
        \caption{Strong Scaling of AF3 with a fixed batch size. As GPU count increases, execution time differences narrow as communication overheads increasingly dominate. }
        \label{fig:strong-scaling}
    \end{minipage}\hfill
    \begin{minipage}{0.47\columnwidth}
        \centering
        \includegraphics[width=\textwidth]{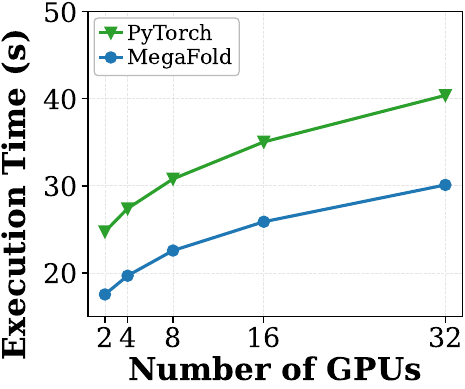}
        \caption{Weak Scaling of AF3 with batch size proportional to GPU count. \name maintains lower execution time than the PyTorch baselines as GPU count increases.}
        \label{fig:weak-scaling}
    \end{minipage}
    \label{fig:strong-weak-scaling}
\end{figure}

\noindent \textbf{Strong scaling.} \fref{fig:strong-scaling} shows strong scaling results when training AF3 with a fixed global batch size of 32 while increasing the number of GPUs from 2 to 32. \name consistently outperforms the baseline across all configurations, achieving a speedup of 1.44$\times$ on 2 GPUs and 1.34$\times$ on 32 GPUs, with an average improvement of 1.38$\times$. The speedups marginally decrease with more GPUs due to increased communication overhead, which indicates the \name's gains are preserved as training becomes more distributed. 
 
\noindent \textbf{Weak scaling.} \fref{fig:weak-scaling} presents weak scaling results, where the batch size scales proportionally with the number of GPUs. \name maintains consistent speedups across all configurations, delivering an average speedup of 1.37$\times$ over \torch baseline from 2 to 32 GPUs. These results demonstrate that \name's sharding and communication strategy scales effectively with increasing problem size.

Together, the strong and weak scaling results show that \name sustains its performance benefits as GPU count increases. This confirms that \sdp enables scalable training over quadratic pairwise representations without introducing prohibitive communication overheads. 

\subsection{Analysis}
\label{subsec:analysis}

\subsubsection{Model Utilization and Scaling Behavior} 
\begin{figure}[t]
    \centering
    \includegraphics[width=0.98\linewidth]{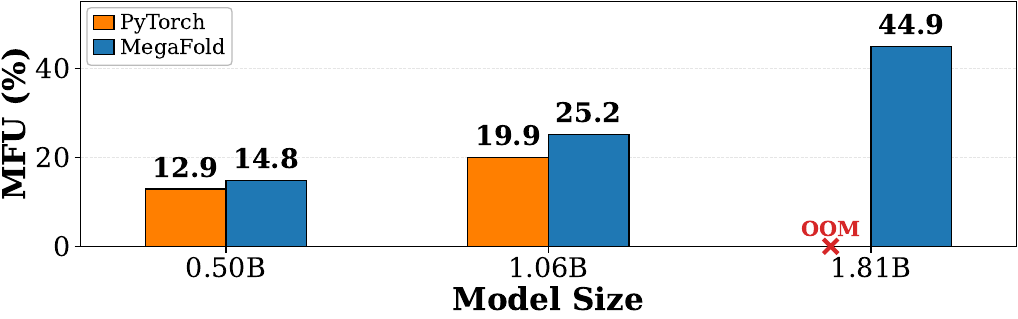}
    \caption{Model FLOPs utilization (MFU) during AF3 training across increasing model sizes. \name consistently improves MFU over the PyTorch baseline and remains trainable at larger model scales, while the baseline runs into OOM. }
    \label{fig:model-scaling}
\end{figure}

\fref{fig:model-scaling} reports model FLOPs utilization (MFU) when training AF3-style models at increasing parameter scales. \name consistently achieves higher MFU than PyTorch baseline across all evaluated configurations. At smaller model size (0.5B), utilization is limited by fixed overheads such as kernel launches and data movement, though \name already provides a modest improvement. As model size increases to 1.06B, \name sustains higher utilization (25.2\% vs. 19.9\%), indicating more effective use of GPU compute. At 1.81B, the PyTorch baseline runs into OOMs, while \name continues to scale and reaches 44.9\% MFU. This shows that \name's utilization gains are enabled by reduced activation memory pressure and kernel fragmentation, allowing training to proceed at scales that are otherwise infeasible. Overall, \name shifts AF training toward a more compute-efficient regime as model size increases, while remaining bounded by memory and communication constraints inherent to 3D attention.  

\subsubsection{Operation-Level Breakdown}
\label{subsubsec:analysis-single-device}
\begin{figure}[!ht]
    \centering
    \includegraphics[width=\linewidth]{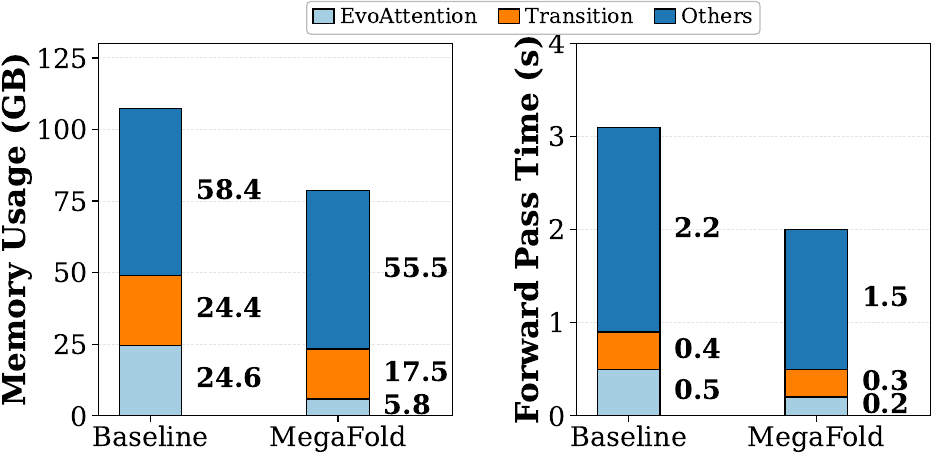}
    \caption{Operation-level breakdown of peak activation memory (left) and forward execution time (right) for AF3 training. \name reduces memory consumption and execution time across \evo, transition layers, and other operators.}
    \label{fig:breakdown-mem-speed}
\end{figure}

We analyze \name's impact at the operator level by breaking down GPU memory consumption and execution time across major operator groups in AF3, as shown in \fref{fig:breakdown-mem-speed}. We group operators into \evo, transition layers, and remaining operators to isolate the effects of \name's kernel-level optimizations. 
\name significantly reduces the memory footprint of the dominant operators in AF3. As shown in \fref{fig:breakdown-mem-speed} (left), the largest reduction occurs in \evo (4.24$\times$), achieved by avoiding materialization of cubic attention maps in HBM through \evoformer. In addition, \name reduces the memory consumption of transition layers (1.39$\times$) and other operators (1.05$\times$) by eliminating redundant intermediate activations via \fusion. These results confirm that \name's kernel level optimizations directly target the primary sources of activation memory pressure identified in \sref{para:memory-heavy-layers}.
\name also substantially reduces the execution time for the dominant operators in AF3. As shown in \fref{fig:breakdown-mem-speed} (right), \evo observes the largest relative speedup (2.5$\times$), which is from reduction of global memory traffic and improved data locality from the \evoformer kernel. Transition layers are accelerated by 1.3$\times$ due to \fusion, which reduces kernel launch overhead and immediate memory access through operator fusion. The execution time of remaining operators is also reduced (1.46$\times$), which contributes to overall performance gains. Together, these results demonstrate that \name's kernel level optimizations effectively improve execution efficiency at the operator granularity. 

\subsubsection{Component Contribution} 
\begin{figure}[!ht]
    \centering
    \includegraphics[width=\linewidth]{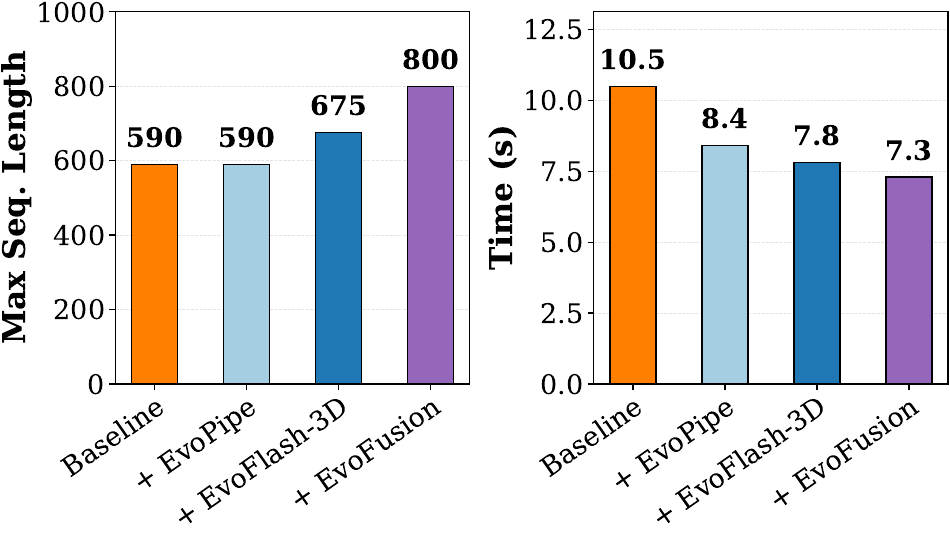}
    \caption{Impact of \name's optimizations on AF3 training. Left: maximum trainable sequence length before OOM. Right:  per-iteration training time. }
    \label{fig:optimization-ablation}
\end{figure}

\fref{fig:optimization-ablation} shows the contribution of individual optimizations on an NVIDIA H200. \data does not increase memory consumption, as its cache is generated offline, while \evoformer and \fusion both contribute significantly to extending the maximum trainable sequence length. In terms of execution time, \name reduces per-iteration time by 1.44$\times$ over PyTorch baseline through a combination of kernel-level and system-level optimizations, including \evoformer, \fusion, and \data. Each component targets a different source of inefficiency in AF3 training, and together they improve performance across the entire training pipeline.

\subsubsection{Scaling on different hardware}
\begin{figure}[t]
    \centering
    \begin{minipage}{0.48\columnwidth}
        \centering
        \includegraphics[width=\textwidth]{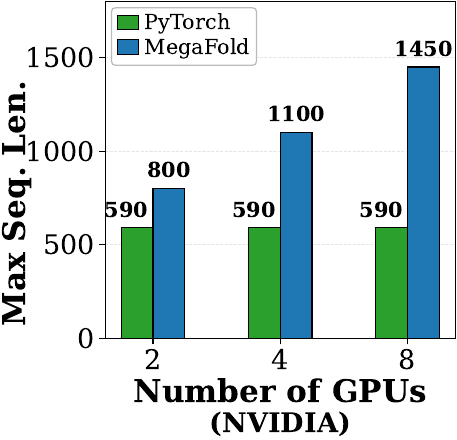}
        \label{fig:nvidia-scaling}
    \end{minipage}\hfill
    \begin{minipage}{0.47\columnwidth}
        \centering
        \includegraphics[width=\textwidth]{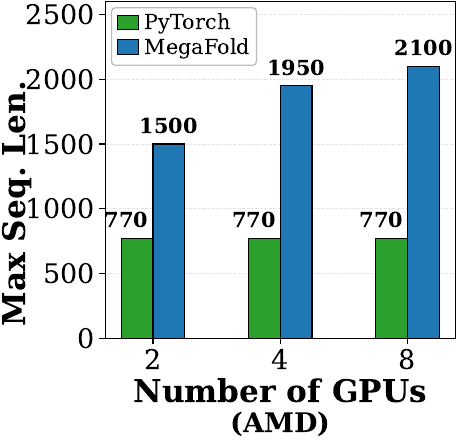}
        \label{fig:amd-scaling}
    \end{minipage}
    \caption{AF3 trainability across various hardware vendors: NVIDIA H200-141GB (left) and AMD MI325x-256GB (right) with a fixed batch size. As GPU count increases, \name scales to training on longer sequence lengths compared to the PyTorch baseline.}
    \label{fig:multi-vendor-scaling}
\end{figure}

~\fref{fig:multi-vendor-scaling}
shows the maximum trainable sequence length on various NVIDIA H200-141GB and AMD MI325x-256GB GPU counts. \name outperforms the PyTorch baseline in all tested configurations, enabling up to a 2.45$\times$ \& 2.72$\times$ increase in input sequence lengths on NVIDIA and AMD hardware, respectively. 
Through a targeted set of optimizations that are designed to be cross-platform, with custom kernels written in Triton,
\name achieves consistent gains across multiple hardware vendors.

\section{Related Work}
\label{sec:related-work}

\noindent
\textbf{Automated fusion and compilation systems.} 
Mirage~\cite{wu2024mirage}, TensorRT~\cite{tensorrt}, OpenVINO~\cite{openvino}, AITemplate~\cite{aitemplate}, TVM~\cite{chen2018tvm}, and IPEX~\cite{ipexllm} explore automated fusion and code-generation for deep neural networks.
However, these systems primarily target inference workloads and do not support the backward pass required for training. NVFuser~\cite{nvfuser} provides automatic kernel fusion for training, but operates only on custom CUDA kernels and assumes relatively regular operator structures. 

\noindent \textbf{Custom kernel libraries.} 
Liger~\cite{liger}, TransformerEngine~\cite{transformerengine}, and xFormers~\cite{xformers} provide highly optimized primitives for transformer models.
However, these libraries do not support the novel mixed-dimensional operators introduced in AF3, such as 3D EvoAttention. 
Composable-kernel (CK)~\cite{composable-kernel} supports training optimizations on AMD GPUs but requires rewriting complex operators in HIP C++ and does not provide out-of-the-box support for AF3-style operators. 

\noindent \textbf{AlphaFold-specific system optimizations.} 
Prior work has proposed system optimizations tailored to AlphaFold models. FastFold~\cite{cheng2023fastfoldreducingalphafoldtraining} introduces Dynamic Axial Parallelism (DAP) and kernel optimizations for the Evoformer module in AlphaFold2~\cite{af2}. ScaleFold~\cite{rajbhandari2020zeromemoryoptimizationstraining} further improves AF2 training with data loading optimizations and specialized kernels for layer normalization and attention. However, these systems target AlphaFold2 and are not directly applicable to AF3, which introduces substantial architectural changes, including structured diffusion, and 3D attention over 2D pairwise representations. In addition, some AF-specific systems are not open-sourced or are limited to NVIDIA hardware.

\section{Conclusion}
\label{sec:conclusion}

AlphaFold3-style models introduce 3D attention and mixed-dimensional operators that fundamentally differ from transformer models, making existing LLM-based optimizations ineffective. \name addresses these challenges through cross-stack co-design, combining memory-efficient 3D attention kernels, communication-efficient sharding for pairwise representations, fused operator restructuring, and determinism-aware host-device pipeline. Our evaluation shows that \name enables longer trainable sequences and faster training across both NVIDIA and AMD GPUs. These results demonstrate the need for system design tailored to emerging scientific models beyond conventional transformer-centric optimizations.

\section*{Acknowledgments}

We sincerely appreciate the anonymous reviewers. Their insightful feedback helps significantly improve the quality of the paper. This research was supported by the National Science Foundation (NSF) under Grant No. 2441601 and 2308699. The work utilized the Delta and DeltaAI system at the National Center for Supercomputing Applications (NCSA) and Jetstream2 at Indiana University through allocation CIS240055 from the Advanced Cyberinfrastructure Coordination Ecosystem: Services \& Support (ACCESS) program, which is supported by National Science Foundation grants \#2138259, \#2138286, \#2138307, \#2137603, and \#2138296. The Delta advanced computing resource is a collaborative effort between the University of Illinois Urbana-Champaign and NCSA, supported by the NSF (award OAC 2005572) and the State of Illinois. This research also used resources of the National Energy Research Scientific Computing Center, a DOE Office of Science User Facility supported by the Office of Science of the U.S. Department of Energy (DOE) under Contract No. DE-AC02-05CH11231 using AI4Sci@NERSC award NERSC DDR-ERCAP0036206 awarded to AM. UIUC SSAIL Lab is supported by research funding and gifts from IBM, Google, Amazon, and AMD, including the Google ML and Systems Junior Faculty Award.

\nocite{*}
\bibliography{_bibliography}

\appendix

\subsection{Triton}
\label{app:background}

Triton is an embedded domain-specific-language in python that allows users to program GPUs with code-generation support for diverse hardware (such as NVIDIA and AMD GPUs), removing the need to rewrite kernels in separate vendor specific DSLs such as CUDA and ROCm. Triton operates on the thread-block level, allowing users to specify how work should decompose across thread-blocks. The Triton compiler then makes a decision on how work is decomposed within the threads constituent to a block. All our code-listings are written in this perspective, detailing the decomposition of work across thread-blocks. 

\subsection{Computations of  EvoAttention \& EvoFlash-3D Backwards Pass}
\label{app:evoattention}

To complement~\sref{subsec:fused-attention}, we illustrate the computations involved in the forward and backward pass of EvoAttention in table~\ref{tab:evo-attention} as well as a code-listing for the backwards pass of EvoFlash-3D in ~\ref{alg:fused-attn-bwd}, detailing its computations next. 

The backwards pass operates in two stages, in the first stage, the \texttt{dV} \& \texttt{dK} tensors are computed (lines 31-47). We schedule each thread-block to compute a chunk of rows in each tensor by fixing a chunk of keys \& values and iterating through the queries. First, each thread-block zero-initializes its local answers of \texttt{dV} \& \texttt{dK} (line 33) and retrieves a fixed chunk of keys and values (line 34-35). Then, the inner-loop (lines 36-46) iterates over the queries and: (1) extracts the correct logsumexp in \texttt{m} and \texttt{D=dO$^T$O} (line 38). The tensor \texttt{D} is precomputed prior to launching the backward pass. (2) computes the transpose of the attention-logits, \texttt{pT} (line 40-42). (3) Multiplies the transposed attention-logits with the incoming gradients to compute \texttt{dV} (line 43). (4) Computes \texttt{dp}-transposed by multiplying the values and the incoming gradients (line 44). (5) Computes \texttt{dS}-transposed (line 45). (6) Finally, computes \texttt{dK} by multiplying \texttt{dS}-transposed by the queries \texttt{q} (line 46).

In the second stage, the \texttt{dQ} and \texttt{dBias} tensors are computed (lines 49-65). In contrast with the first stage, we schedule a thread-block to compute a chunk of rows in each tensor by fixing a chunk of queries and iterating through the values and keys. The preamble prior to the loop is similar to stage one. The inner-loop (lines 55-64) first computes a chunk of the \texttt{s}, \texttt{p}, \texttt{dp}, and \texttt{ds} tensors. Then, the \texttt{dBias} gradients are computed by reducing the local \texttt{ds} chunk through an \texttt{atomic\_add} to the corresponding \texttt{dBias} memory location; doing so mitigates the need to hold partially reduced values in scratchpad memories and avoids impacting the occupancy of our kernel. Finally, the \texttt{dQ}-chunk is computed by multiplying \texttt{ds} by the keys, \texttt{k}.

\definecolor{codered}{rgb}{0.6,0,0}
\definecolor{codegreen}{rgb}{0,0.6,0}
\definecolor{codeblue}{rgb}{0,0,0.6}
\definecolor{codepurple}{rgb}{0.58,0,0.82}
\definecolor{codebg}{rgb}{0.97,0.97,0.97}

\lstdefinestyle{fused-attn-style}{
    language=Python,
    frame=single,
    framerule=0pt,
    backgroundcolor=\color{codebg},   
    commentstyle=\color{codepurple},
    keywordstyle=\color{codeblue},
    emphstyle=\ttfamily\color{codered},
    stringstyle=\color{codegreen},
    basicstyle=\ttfamily\linespread{0.85}\footnotesize,
    breakatwhitespace=false,         
    breaklines=true,                 
    captionpos=b,                    
    keepspaces=true,                 
    numbers=left,                    
    numbersep=-4pt,                  
    showspaces=false,                
    showstringspaces=false,
    showtabs=false,                  
    tabsize=2,
    morekeywords={}
}

\lstset{
  mathescape,         
  literate={->}{$\rightarrow$}{2}
           {ε}{$\varepsilon$}{1}
}

\begin{lstlisting}[style={fused-attn-style}, 
    label={alg:fused-attn-bwd},
    caption={EvoFlash-3D backwards pass.},
    float=!t]
  # Omit indexing on batch/head/hid. dim for brevity
  def fused_attn_bwd(dO,Q,K,V,bias,l,m,D,O):
    D = preprocess(dO,O)
    dk, dv = dv_dk(dO,Q,K,V,bias,l,m,D)
    dBias = zeros(bias.shape)
    dq = dq_dbias(dO,Q,K,V,bias,l,m,D,dBias)
    return dq, dk, dv, dbias
  # Stage 1, compute: (1)dv, (2)dP, (3)dS, (4)dk
  def dv_dk(dO,Q,K,V,bias,l,m,D):
    j = tl.program_id(0)
    dk = dv = zeros((bx, d))
    kj, vj = k[j*bx:(j+1)*bx, :], \
          v[j*bx:(j+1)*bx, :]
    for i in range(N/by):
      qiT = q[i*by:(i+1)*by, :].T
      mi, di = m[i*by:(i+1)*by], D[i*by:(i+1)*by]
      dOi = dO[i*by:(i+1)*by, :]
      sT = tl.matmul(kj, qiT) + \
              bias[j*bx:(j+1)*bx,i*by:(i+1)*by]
      pT = tl.exp(sT-mi)
      dv += tl.matmul(pT, dOi)
      dpT = tl.matmul(v, dOi.T)
      dST = pT * (dpT - di)
      dk += tl.matmul(dST, qiT.T)
    return dk, dv
  # Stage 2, compute: (1)dS, (2)dbias, (3)dq
  def dq_dbias(dO,Q,K,V,bias,l,m,D,dBias):
    i = tl.program_id(0)
    dq = zeros((by, d))
    qi = Q[i*by:(i+1)*by, :]
    doi = dO[i*by:(i+1)*by, :]
    mi, di = m[i*by:(i+1)*by], D[i*by:(i+1)*by]
    for j in range(N/bx):
      kjT, vjT = K[j*bx:(j+1)bx, :].T, \
                  V[j*bx:(j+1)*bx, :].T
      s = tl.matmul(qi, kjT) + \
              bias[i*by:(i+1)*by,j*bx:(j+1)*bx]
      p = tl.exp(s-mi)
      dp = tl.matmul(doi, vjT)
      ds = p*(dp-di)
      tl.atomic_add(dbias, ds)
      dq += tl.matmul(ds, kjT.T)
    return dq\end{lstlisting}


\begin{table}[!htbp]
\centering
\caption{EvoAttention's forward and backward passes.}
\label{tab:evo-attention}
\begin{tabular}{c|c}
\toprule
\small
\textbf{Forward Pass} & \textbf{Backward Pass (given $\texttt{dO} = \frac{\partial \mathcal{L}}{\partial \texttt{O}}$)} \\
\midrule
$\displaystyle
\begin{aligned}
    \texttt{S} &= \texttt{QK$^\texttt{T}$} \\
      &\quad + \texttt{Bias$_{\texttt{bcast}}$} \\
    \texttt{P} &= \texttt{Softmax(S)} \\
    \texttt{O} &= \texttt{PV}
\end{aligned}
$
&
$\displaystyle
\begin{aligned}
    \texttt{dV} &= \texttt{P$^\texttt{T}$} \texttt{dO} \\
    \texttt{dP} &= \texttt{dOV$^T$} \\
    \texttt{dS} &= \texttt{P} \odot (\texttt{dP} \\
      &\quad - \underbrace{\texttt{rowsum(dO$^T$O)}}_{\texttt{D}})\\
    \texttt{dK} &= \texttt{Q$^T$dS}\\
    \texttt{dBias} &= \!\sum_{\texttt{bcast}}\!\texttt{dS} \\
    \texttt{dQ} &= \texttt{dS K}
\end{aligned}
$
\\
\bottomrule
\end{tabular}
\end{table}

\subsection{Computations of LayerNorm \& Linear Layer}
\label{app:ln-ops}

To complement~\sref{subsec:fused-small-ops}, we illustrate the computations involved in the forward and backward pass of a layer-norm \& linear-layer in table~\ref{tab:fused-small-ops}. The \texttt{c$_\text{1}$} and \texttt{c$_\text{2}$} variables are intermediate values that are used to compute the gradients of the input tokens. We additionally detail the forward and backward passes in listings~\ref{alg:fused-ln-linear} and ~\ref{alg:fused-ln-linear-bwd} respectively. Since the forward pass is detailed in~\sref{subsec:fused-small-ops}, we detail only its backwards pass next.

Our backward pass launches two kernels (lines 34-63 in listing~\ref{alg:fused-ln-linear}). The first kernel computes the gradients of the linear-layer weights, the weights of the LayerNorm parameters, as well as extra \texttt{c1}, \texttt{c2}, and \texttt{wdy} data, intermediate variables required to compute the gradients of the \texttt{token$_\texttt{p}$} tensor (the mathematical formulae for these intermediate values are described in table~\ref{tab:fused-small-ops}). The second kernel then consumes these data to finally compute the gradients of the \texttt{token$_\texttt{p}$} tensor. 

The first kernel first reads the input \texttt{token} tensor and re-normalizes the data by applying the LayerNorm weights and bias (lines 36-40), and then zero-initializes a \texttt{[b$_{\texttt{m}}$, b$_{\texttt{k}}$]}-sized tile where a local tile of \texttt{do$_\texttt{1}$} will be computed. An inner-loop (lines 42-48) then traverses across the linear-layer's hidden dimension and computes: (1) a tile of \texttt{dW$_\texttt{1}$}, atomically aggregating this to main-memory, where the overall result of \texttt{dW$_\texttt{1}$} is stored; (2) updates the local tile of \texttt{dO$_\texttt{1}$}. After the inner-loop terminates, the local tile of \texttt{dO$_\texttt{1}$} is fully computed and is used to compute \texttt{wdy} (line 49), which is then used to compute \texttt{c$_\texttt{1}$} \& \texttt{c$_\texttt{2}$} by atomically aggregating local values of \texttt{c$_\texttt{1}$} \& \texttt{c$_\texttt{2}$} to main-memory, where their overall results are stored (lines 50-51). Finally, the LayerNorm weights \texttt{dw} and \texttt{db}  are computed (lines 52-53) and the \texttt{wdy} tensor is returned to be consumed by the second kernel in the backwards pass. In the second kernel (lines 56-63), each thread-block computes a chunk of rows of \texttt{dtoken$_\texttt{p}$} by first reading the original input tokens from the \texttt{token} tensor, normalizing the loaded input and finally using the \texttt{c$_\texttt{1}$}, \texttt{c$_\texttt{2}$}, \texttt{wdy} \& normalized inputs to compute \texttt{dtoken$_\texttt{p}$}.

\begin{table*}[!htbp]
\centering
\caption{\fusion for Common AlphaFold Operators}
\label{tab:fused-small-ops}
\renewcommand{\arraystretch}{1.8} 

\begin{tabular}{c|c}
\toprule
\textbf{Forward Pass} & \textbf{Backward Pass (given $\texttt{dO} = \frac{\partial \mathcal{L}}{\partial \texttt{O$_\texttt{2}$}}$)} \\
\midrule
$\displaystyle 
\begin{aligned}
\texttt{O$_\texttt{1}$} &= \underbrace{\frac{\texttt{token$_\text{p}$} - \mathbb{E}[\texttt{token$_\text{p}$}]}{\sqrt{\texttt{Var}(\texttt{token$_\text{p}$}) + \epsilon}}}_{\hat{\texttt{token$_\text{p}$}}} \odot \texttt{w + b}\\
\texttt{O$_\texttt{2}$} &= \texttt{O$_\texttt{1}$W$_\texttt{1}$} \\
\end{aligned}
$
&
$\displaystyle 
\begin{aligned}
    \texttt{dW$_\texttt{1}$} &= \texttt{O$_\texttt{1}$$^\texttt{T}$dO$_\texttt{2}$} \\
    \texttt{dO$_\texttt{1}$} &= \texttt{dO$_\texttt{2}$W$_\texttt{1}$$^\texttt{T}$} \\
    \texttt{dw} &= \texttt{dO$_\texttt{1}$} \odot \hat{\texttt{token$_\text{p}$}} \\
    \texttt{db} &= \sum_{\texttt{dimension=0}} \texttt{dO$_\texttt{1}$} \\
    \texttt{d}\texttt{token$_\text{p}$} &= \frac{1}{\texttt{Var}(\texttt{token$_{\text{p}}$})}\bigg(\texttt{dO$_\texttt{1}\odot$w} - \texttt{c$_\texttt{1}$} \odot \hat{\texttt{token$_\text{p}$}} - \texttt{c$_\texttt{2}$} \bigg) \\
    \texttt{c$_\texttt{1}$} &= \frac{1}{\texttt{N$_{\texttt{token}}$}} \sum_{\texttt{dimension}=1} \hat{\texttt{token$_\text{p}$}} \odot (\texttt{dO$_\texttt{1}$} \odot \texttt{w}) \\
    \texttt{c$_\texttt{2}$} &= \frac{1}{\texttt{N$_{\texttt{token}}$}} \sum_{\texttt{dimension}=1} \texttt{dO$_\texttt{1}$}\odot\texttt{w}
\end{aligned}
$
\\
\bottomrule
\end{tabular}
\end{table*}

\definecolor{codered}{rgb}{0.6,0,0}
\definecolor{codegreen}{rgb}{0,0.6,0}
\definecolor{codeblue}{rgb}{0,0,0.6}
\definecolor{codepurple}{rgb}{0.58,0,0.82}
\definecolor{codebg}{rgb}{0.97,0.97,0.97}

\lstdefinestyle{fused-attn-style}{
    language=Python,
    frame=single,
    framerule=0pt,
    backgroundcolor=\color{codebg},   
    commentstyle=\color{codepurple},
    keywordstyle=\color{codeblue},
    emphstyle=\ttfamily\color{codered},
    stringstyle=\color{codegreen},
    basicstyle=\ttfamily\linespread{0.85}\footnotesize,
    breakatwhitespace=false,         
    breaklines=true,                 
    captionpos=b,                    
    keepspaces=true,                 
    numbers=left,                    
    numbersep=-4pt,                  
    showspaces=false,                
    showstringspaces=false,
    showtabs=false,                  
    tabsize=2,
    morekeywords={}
}

\lstset{
  mathescape,         
  literate={->}{$\rightarrow$}{2}
           {ε}{$\varepsilon$}{1}
}

\begin{lstlisting}[style={fused-attn-style}, 
    label={alg:fused-ln-linear},
    caption={Fused Transition layer forward pass.},
    float=!ht]
  def transition_fwd(token, eps, w, b, W$_1$, W$_2$):
    mean, rstd = mean_var(token, eps)
    o$_1$ = fused_ln_linear_fwd(token, eps, mean, \
                        rstd, w, b, W$_1$)
    o$_2$ = fused_swiglu_fwd(o$_1$) 
    return o = matmul(o$_2$, W$_2$)
  def transition_bwd(dO, w, b, W$_1$, W$_2$):
    ... = fused_swiglu_bwd(...)
    do$_1$ = ... # Omit computing earlier gradients 
    c1 = c2 = zero_init((M,1))
    dW$_1$, dw, db, c1, c2, wdy = \
                dW$_1$_dw_db_dc1_dc2(...)
    dtokenp = dtokenp(...)
  def mean_var(token, eps):
    row, k = token[tl.program_id(0)], \
                token.shape(-1)
    mean = tl.sum(row, axis=-1) / k
    var = tl.sum((row-mean)**2, axis=0) / k
    rstd = 1/tl.sqrt(var+eps)
    return mean, rstd
  def fused_ln_linear_fwd(token, eps, mean, rstd, w, b, W$_1$):
    m, n = swizzle(tl.program_id(0))
    o2 = zeros((bm, bn))
    for k in range(K/bk); # K is the hid. dimension
      # Load (bm, bk)-sized chunk
      x = token[m*bm:(m+1)*bm, k*bk:(k+1)*bk] 
      wi, bi = w[k*bk:(k+1)*bk], b[k*bk:(k+1)*bk]
      # Load (bk, bn)-sized chunk
      y = W$_1$[k*bk:(k+1)*bk,n*bn:(n+1)*bn]
      $\hat{x}$ = ((x - mean) * rstd) * wi + bi
      o2 += tl.matmul($\hat{x}$, y) # (bm,bk)x(bk,bn)
    return o2\end{lstlisting}
\definecolor{codered}{rgb}{0.6,0,0}
\definecolor{codegreen}{rgb}{0,0.6,0}
\definecolor{codeblue}{rgb}{0,0,0.6}
\definecolor{codepurple}{rgb}{0.58,0,0.82}
\definecolor{codebg}{rgb}{0.97,0.97,0.97}

\lstdefinestyle{fused-attn-style}{
    language=Python,
    frame=single,
    framerule=0pt,
    backgroundcolor=\color{codebg},   
    commentstyle=\color{codepurple},
    keywordstyle=\color{codeblue},
    emphstyle=\ttfamily\color{codered},
    stringstyle=\color{codegreen},
    basicstyle=\ttfamily\linespread{0.85}\footnotesize,
    breakatwhitespace=false,         
    breaklines=true,                 
    captionpos=b,                    
    keepspaces=true,                 
    numbers=left,                    
    numbersep=-4pt,                  
    showspaces=false,                
    showstringspaces=false,
    showtabs=false,                  
    tabsize=2,
    morekeywords={}
}

\lstset{
  mathescape,         
  literate={->}{$\rightarrow$}{2}
           {ε}{$\varepsilon$}{1}
}

\begin{lstlisting}[style={fused-attn-style}, 
    label={alg:fused-ln-linear-bwd},
    caption={Fused Transition layer backward pass.},
    float=!ht]
  def transition_bwd(dO, w, b, W$_1$, W$_2$):
    ... = fused_swiglu_bwd(...)
    do$_1$ = ... # Omit computing earlier gradients 
    c1 = c2 = zero_init((M,1))
    dW$_1$, dw, db, c1, c2, wdy = \
                dW$_1$_dw_db_dc1_dc2(...)
    dtokenp = dtokenp(...)
  # computes dw, db, c1 and c2
  def dW$_1$_dw_db_dc1_dc2(dO2, dW$_1$, token, mean, rstd, w, b, W$_1$, c1_ptr, c2_ptr):
    m, k = swizzle(tl.program_id(0))
    x = token[m*bm:(m+1)*bm, k*bk:(k+1)*bk]
    meani, rstdi = mean[m*bm:(m+1)*bm], \
                    rstd[m*bm:(m+1)*bm]
    xhat = (x - meani) * rstdi
    yiT = tl.trans(xhat*w + b) # Layer-norm output
    dO1 = zero_init((bm, bk))
    for n in range(N/bn): # N is the ffn hid. dim 
      dO2 = dO2[m*bm:(m+1)*bm,n*bn:(n+1)*bn]
      dW1 = matmul(yiT, dO2)
      tl.atomic_add(dW$_1$[k*bk:(k+1)*bk, \
                    n*bn:(n+1)*bn], dW1)
      W$_1$T = W$_1$[k*bk:(k+1)*bk,n*bn:(n+1)*bn].T
      dO1 += matmul(dO2, W$_1$T)
    wdy = dO1 * w
    tl.atomic_add(c1_ptr, sum(xhat * wdy, dim=1)/K)
    tl.atomic_add(c2_ptr, sum(wdy, dim=1)/K)
    tl.atomic_add(dw_ptr, sum(dO1 * xhat, dim=0))
    tl.atomic_add(db_ptr, sum(dO1, dim=0))
    return wdy
  # computes dtokenp
  def dtokenp(c1, c2, wdy, token, mean, rstd):
    i = tl.program_id(0) 
    xi = token[i] # Block operates on a row. 
    c1i, c2i, meani, rstdi, wdyi = c1[i], c2[i], \
            mean[i], rstd[i], wdy[i]
    xhati = (xi-meani) * rstdi
    dtokenpi = (wdyi - (c1i * xhati + c2i)) * rstdi
    return dtokenpi\end{lstlisting}

\subsection{EvoPipe}
\label{app:caching}

To complement~\sref{subsec:data-loading}, we illustrate the default and optimized training loop prior/post application of EvoPipe's determinism-aware host-device pipeline in listing~\ref{alg:training-loop}.

\definecolor{codered}{rgb}{0.6,0,0}
\definecolor{codegreen}{rgb}{0,0.6,0}
\definecolor{codeblue}{rgb}{0,0,0.6}
\definecolor{codepurple}{rgb}{0.58,0,0.82}
\definecolor{codebg}{rgb}{0.97,0.97,0.97}

\lstdefinestyle{fused-attn-style}{
    language=Python,
    frame=single,
    framerule=0pt,
    backgroundcolor=\color{codebg},   
    commentstyle=\color{codepurple},
    keywordstyle=\color{codeblue},
    emphstyle=\ttfamily\color{codered},
    stringstyle=\color{codegreen},
    basicstyle=\ttfamily\linespread{0.85}\footnotesize,
    breakatwhitespace=false,         
    breaklines=true,                 
    captionpos=b,                    
    keepspaces=true,                 
    numbers=left,                    
    numbersep=-4pt,                  
    showspaces=false,                
    showstringspaces=false,
    showtabs=false,                  
    tabsize=2,
    morekeywords={}
}

\lstset{
  mathescape,         
  literate={->}{$\rightarrow$}{2}
           {ε}{$\varepsilon$}{1}
}

\begin{lstlisting}[style={fused-attn-style}, 
    label={alg:training-loop},
    caption={Main training loop after application of EvoPipe.},
    float=!ht]
  def generate_cache(data):
    cache = {}
    for inp in data:
      cache[inp] = (atom_features(inp),\
                        msa_features(inp))
    return cache
  def train_loop(data,af,opt,loss_fn):
    """Main training loop.
    data: input data-source.
    af: the af-model to train.
    """
    for inp, out in data:
      # Extract atom feature info  
      atom_feat = atom_features(inp)
      # Extract msa feature info
      msa_feat = msa_features(inp)
      p_atom_feat,p_msa_feat = preprocess(atom_feat, 
                                msa_feat) 
      tokenize_repr = tokenize(p_atom_feat,
                        p_msa_feat)
      output = loss_fn(af(tokenize_repr), output)
      output.backward()
      opt.step()
  def cached_train(data,af,opt,loss_fn, cache):
    """Main training loop.
    data: input data-source.
    af: the af-model to train.
    """
    for inp, out in data:
      atom_feat, msa_feat = cache[inp]
      p_atom_feat,p_msa_feat = preprocess(atom_feat, 
                                msa_feat) 
      tokenize_repr = tokenize(p_atom_feat,
                        p_msa_feat)
      output = loss_fn(af(tokenize_repr), output)
      output.backward()
      opt.step()
  # Stage 1: Populate cache.
  cache = generate_cache(data)
  # Stage 2: train. 
  cached_train(data, af, opt, loss, cache)\end{lstlisting}

\subsection{Activation Checkpointing Setup}
\label{app:ac-setup}
Given the large activations memory produced during AF3 training, activation-checkpointing (AC) must be enabled otherwise it is not possible to train AF3 on moderate sequence lengths. AC is systematically applied to every sub-layer within each layer in AF3 modules. Using the terminology in the original AF3 paper, we apply AC to the following:
\begin{itemize}
    \item MSA module: AC is applied on each OuterProductMean, MSAPairWeightedAveraging, Transition, and PairwiseBlock sub-layer
    \item Template module: AC is applied on each PairwiseBlock sub-layer
    \item Pairformer module: AC is applied on each PairwiseBlock, AttentionPairBias, and Transition sub-layer
    \item Diffusion module: AC is applied on each AttentionPairBias, Transition sub-layer
\end{itemize}

When AC is applied, all intermediate activations within the sub-layer are discarded during the forward pass, and only the input tensors are retained. In the backward pass, since the intermediate activations have been discarded, an additional forward pass is performed using the retained inputs to recompute necessary activations for gradient calculations. AC accordingly makes a trade-off between speed and memory, sacrificing speed for reduced peak memory consumption. On the other hand, our optimizations achieve the best of both worlds: faster training speeds with reduced peak memory consumption.

%
%

\subsection{Fraction of theoretical peak achieved of Individual Optimizations}

\begin{figure}[!ht]
    \centering
    \includegraphics[width=\linewidth]{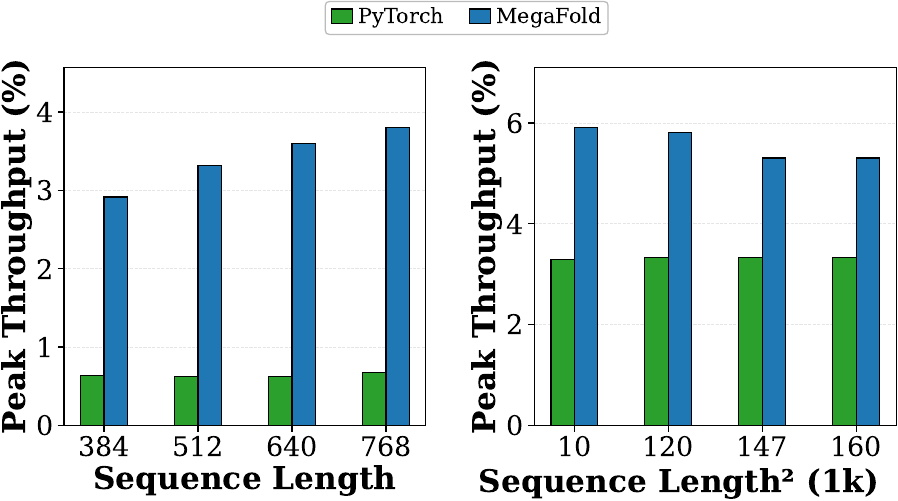}
    \caption{Comparing fraction of theoretical throughput achieved of \name optimizations and the PyTorch baseline. Left and Right are the EvoAttention and Transition operators respectively. \name optimizations and custom kernels consistently outperform PyTorch.}
    \label{fig:peak-throughput-cmp}
\end{figure}

We compare the fraction of theoretical peak achieved when applying \name \evoformer and \fusion optimizations to AF3 operators against PyTorch in~\fref{fig:peak-throughput-cmp}. \name EvoFlash-3D and EvoFused operators achieve upto 4 and 6\% of peak throughput, respectively. Comparatively, PyTorch achieves $<$1\% and 3\% of peak throughput, indicating our optimizations effectively reduce the number of reads/writes to HBM through fusion to accelerate performance. Note that the input sizes to these kernels are relatively small (typical of AF3 models), resulting in a significantly lower achievable throughput compared to larger models like LLMs.

\subsection{Implementation Validation}
\label{app:impl-validation}

\begin{figure}[!ht]
    \centering
    \includegraphics[width=\linewidth]{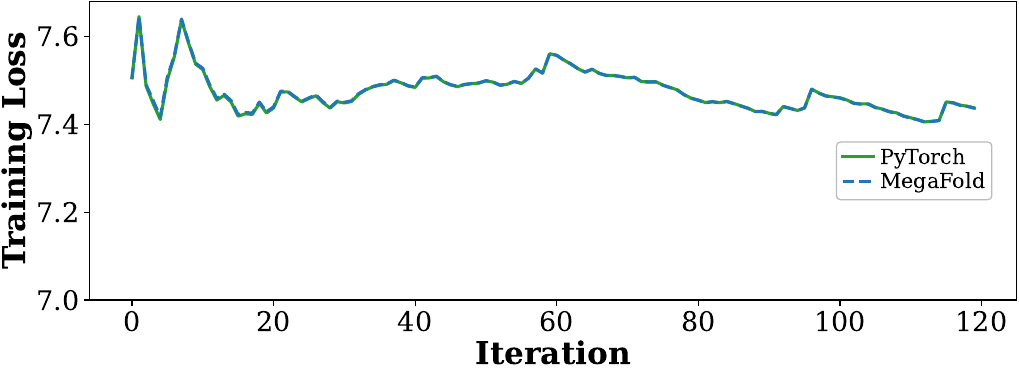}
    \caption{Comparing training loss between \name and PyTorch of AF3 across 120 steps. \name and PyTorch exhibit the same convergence.}
    \label{fig:convergence-test}
\end{figure}

Fused kernels often modify reductions and may change the numerical stability of operators. Therefore, to verify the correctness of our optimizations, we train \name \& a PyTorch implementation of AF3 for 120 iterations with a sequence length of 384 using the Protein Data Bank (the defacto dataset to train AF3), plotting the training loss at each step in \fref{fig:convergence-test}. \name matches the per-step training loss of the baseline, indicating that our optimizations do not sufficiently impact AF3's convergence. 

\subsection{Comparison against Hand-written EvoAttention Kernels}
\label{app:compare-trifast}

\begin{figure}[!ht]
    \centering
    \includegraphics[width=\linewidth]{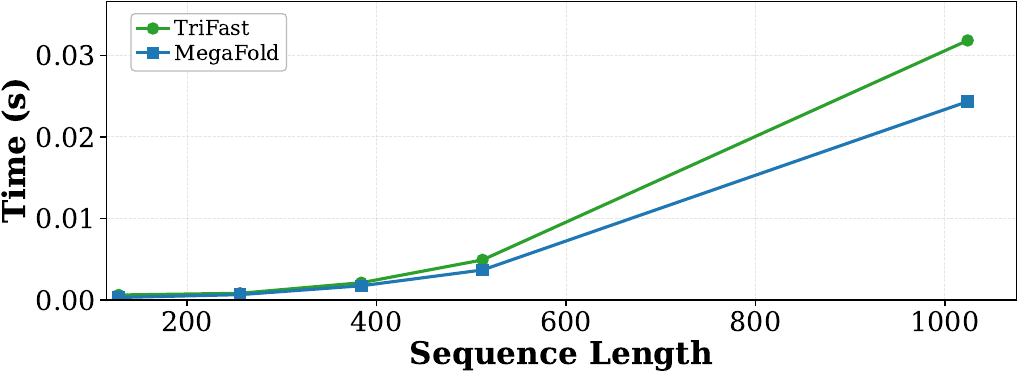}
    \caption{Comparing the runtime performance of TriFast and MegaFold's EvoAttention operators across different sequence lengths. Timing aggregates forward and backwards pass. \name consistently outperforms TriFast across all sequence lengths.}
    \label{fig:megafold-trifast}
\end{figure}

We compare the runtime performance of MegaFold's memory-efficient EvoAttention operator against TriFast~\cite{trifast}, a hand-written EvoAttention implementation heavily used by industry in~\fref{fig:megafold-trifast}. Across all benchmarked sequence lengths, MegaFold's EvoAttention operator - EvoFlash-3D - is consistently 1.2$\times$-2$\times$ faster than TriFast. This speedup is attributed to the backwards pass. TriFast launches three kernels to compute gradients: (1) \texttt{dK \& dV}, (2) \texttt{dQ} and (3) \texttt{dBias}. On the other hand, MegaFold launches two kernels: (1) \texttt{dK \& dV}, and (2) \texttt{dQ \& dBias}, comparatively reducing kernel launch overhead. 

\end{document}